\documentclass[acmsmall,screen,authorversion,nonacm]{acmart} % PREPRINT
\usepackage{xspace}
\usepackage{graphicx}
\graphicspath{{figures/}}
\usepackage{needspace}
\newcommand{\needlines}[1]{\Needspace{#1\baselineskip}}
\usepackage{paralist}
\usepackage{ifthen}
\usepackage[normalem]{ulem} % for \sout
\usepackage{xcolor}

\newboolean{showedits}
\setboolean{showedits}{true} % toggle to show or hide edits
\ifthenelse{\boolean{showedits}}
{
	\newcommand{\del}[1]{\textcolor{red}{\sout{#1}}} % please delete
	\newcommand{\nbe}[3]{
		{\colorbox{#3}{\bfseries\sffamily\scriptsize\textcolor{white}{#1}}}
		{\textcolor{#3}{\sf\small$\blacktriangleright$\textit{#2}$\blacktriangleleft$}}}
}{
	\newcommand{\del}[1]{} % please delete
	
	\newcommand{\nbe}[3]{}
}
\usepackage[most]{tcolorbox}
\ifthenelse{\boolean{showedits}}
{
  \newtcolorbox{inserted}{%
       title=Inserted text:,
       colframe=blue,colback=blue!5!white,
       breakable,
       leftrule=0mm, 
       bottomrule=0mm,
       rightrule=0mm,
       toprule=0mm,
       arc=0mm, outer arc=0mm,
       oversize
  }
  \newtcolorbox{deleted}{%
       title=Deleted text:,
       colframe=red,colback=red!5!white,
       breakable,
       leftrule=0mm, 
       bottomrule=0mm,
       rightrule=0mm,
       toprule=0mm,
       arc=0mm, outer arc=0mm,
       oversize
  }
  \newtcolorbox{refactored}{%
       title=Rewritten text:,
       colframe=blue,colback=red!5!white,
       breakable,
       leftrule=0mm, 
       bottomrule=0mm,
       rightrule=0mm,
       toprule=0mm,
       arc=0mm, outer arc=0mm,
       oversize
  }
}{

}
\newboolean{showcomments}
\setboolean{showcomments}{true}
\newcommand{\id}[1]{$-$Id: scgPaper.tex 32478 2010-04-29 09:11:32Z oscar $-$}

\ifthenelse{\boolean{showcomments}}
{\newcommand{\nbc}[3]{
 {\colorbox{#3}{\bfseries\sffamily\scriptsize\textcolor{white}{#1}}}
 {\textcolor{#3}{\sf\small$\blacktriangleright$\textit{#2}$\blacktriangleleft$}}}
 }
{\newcommand{\nbc}[3]{}
 }

\newboolean{isblinded}
\setboolean{isblinded}{true}
\ifthenelse{\boolean{isblinded}}
{\newcommand\blind[1]{BLINDED\xspace}}
{\newcommand\blind[1]{#1\xspace}}
\newcommand{\ie}{\emph{i.e.},\xspace}
\newcommand{\eg}{\emph{e.g.},\xspace}

\newcommand{\etc}{\emph{etc.}\xspace}
\newcommand{\sep}{\mbox{$\gg$}}
\usepackage[english]{babel}
\usepackage{listings}
\lstdefinelanguage{Smalltalk}{
  morestring=[d]',
  morecomment=[s]{"}{"},
  alsoletter={\#:},
  escapechar={!},
  literate=
    {BANG}{!}1
    {UNDERSCORE}{\_}1
    {_}{{$\leftarrow$}}1
    {>>>}{{\sep}}1
    {^}{{$\uparrow$}}1
    {~}{{$\sim$}}1
    {-}{{\sf -\hspace{-0.13em}-}}1  % the goal is to make - the same width as +
    {+}{\raisebox{0.08ex}{+}}1		% and to raise + off the baseline to match -
    {-->}{{\quad$\longrightarrow$\quad}}3
	, % Don't forget the comma at the end!
  tabsize=4
}[keywords,comments,strings]

\definecolor{source}{gray}{0.95}

\newcommand{\st}{\lstinline[mathescape=false,backgroundcolor=\color{white},basicstyle={\sffamily\upshape}]}
\newcommand{\lst}[1]{{\textsf{\textup{#1}}}}
\lstnewenvironment{code}{%
	\lstset{%
		frame=single,
		framerule=0pt,
		mathescape=false
	}
}{}

\newboolean{preprint}
\setboolean{preprint}{true}
\AtBeginDocument{%
  }
\setcopyright{acmlicensed}
\copyrightyear{2024}
\acmYear{2024}
\acmDOI{XXXXXXX.XXXXXXX}
\acmConference[EuroPLoP '24]{29th European Conference on Pattern Languages of Programs}{July 3--7, 2024}{Kloster Irsee, Germany}
\acmISBN{978-1-4503-XXXX-X/18/06}
\usepackage{caption}
\newcommand{\GT}{\lst{GT}\xspace} % In case we want to display it differently ...
\newcommand{\pattern}[2]{\needlines{10}
\subsection*{#1}\label{pat:#2}}
\newcommand{\patref}[1]{\emph{\nameref{pat:#1}}\xspace}
\newcommand{\patsec}[1]{\noindent\textit{#1.}\xspace}
\begin{document}
\ifthenelse{\boolean{preprint}}{% FOR PREPRINT
\title[Moldable Development Patterns --- preprint]{Moldable Development Patterns}%
\thanks{Preprint of paper workshopped at \href{https://www.europlop.net}{EuroPLoP} 2024, July 3--7, 2024, Kloster Irsee, Germany. To appear (ACM)}%
}{
\title{Moldable Development Patterns}%
}

\author{Oscar Nierstrasz}
\affiliation{%
  \institution{feenk gmbh}
  \city{Wabern}
  \country{Switzerland}}
\email{oscar.nierstrasz@feenk.com}

\author{Tudor G\^irba}
\affiliation{%
  \institution{feenk gmbh}
  \city{Wabern}
  \country{Switzerland}}
\email{tudor.girba@feenk.com}

\begin{abstract}
%\emph{NB: This draft has been formatted single-column with 1.4x spacing for the EuroPLoP conference. This is not the final print version.}

Moldable development supports decision-making by making software systems \emph{explainable}.
This is done by making it cheap to add numerous custom tools to your software, 
%\cp{your software or your software development environment?}
turning it into a live, explorable domain model.
Based on several years of experience of applying moldable development to both open-source and industrial systems, we have identified several mutually supporting patterns to explain how moldable development works in practice.
This paper targets
(i) readers curious to learn about moldable development,
(ii) current users of the \emph{Glamorous Toolkit} moldable IDE
%\cp{these are probably the readers who will benefit most, given the examples and terminology used which is very GT specific.}
wanting to learn best practices, and
(iii) developers interested in applying moldable development using other platforms and technology.
%\dd{Good explanation of the paper's target audience}
\end{abstract}


\maketitle

% ===== Introduction =========================
\section{Introduction: Moldable development in a Nutshell}

%\ws{Takeaways from the workshop:\\
%- Separate solution summary from details. \\
%- List separately related patterns.\\
%- List more negative consequences.\\
%- List known uses explicitly.\\
%- Highlight better the patterns in the paper.\\
%- Perhaps describe an idealized order of the patterns.\\
%- Which patterns are obligatory, resp. optional?\\
%- The Explainable System motivation needs to be clearer.\\
%- The experience should be made clear at the beginning, not the end.\\
%- Explain the difference between a Simple View and a raw view.\\
%- Also: it would have been better to split the patterns into detailed patterns and “patlets” just described in a line or two.
%}

Software systems are rich sources of knowledge for both developers and non-technical stakeholders.
But it can be difficult to extract that knowledge.
The usual artifacts to evaluate software systems are
\begin{inparaenum}[(i)]
\item the source code, and
\item the running system.
\end{inparaenum}
But neither of these lends itself well to answering questions about the system.
%\cp{ignoring requirement specifications, documentation, modeling artifacts, architectural decision records...}
In addition there may exist documentation and other related artifacts, but these are typically incomplete, or out-of-sync with the actual system.
Software analysis tools can help to some extent, but since every system and every problem is different, \emph{it is rare for generic analysis tools to be effective for arbitrary systems.}

When we think of software development, we typically think about the active part.
Of constructing.
Of building new worlds that never existed.
It's an empowering view.
Yet, developers spend most of their time figuring systems out~\cite{Xia18a}.
% Was \cite{Zelk79b}
They do that because they want to learn enough about the system to make a decision.
This is the single largest development expense we have.
So, we should optimize software engineering for decision making.

Moldable development is an approach to constructing software systems that are \emph{explainable}, that is to say, systems that can answer questions about themselves.
%\cp{What is your definition of "explainability"?  Is it the ability of answering questions about the software? but who answers the questions? the tool or the developer interpreting what a tool shows?}
This is achieved by making it inexpensive to create dozens, hundreds or even thousands of custom tools to answers questions about a software system \emph{as these questions arise}.
These custom tools consist of small extensions to the \emph{moldable} tools of the integrated development environment (IDE), such as the object inspector, the code editor, the debugger and the notebook.

Every part of a software system deals with a particular \emph{domain}, whether this be the \emph{business domain} of the overall application, or a more \emph{technical domain}  of the underlying application.
%\ws{JN: Is the business model separate from the display? (cf MVC)  In the back of the Blue book there is a reference to a non-existing book on Building Interactive Applications in Smalltalk. The GoF book was the closest answer.}
The net effect of moldable development is to \emph{expose} the underlying models of these diverse domains (\ie the ``domain models'') 
%\cp{is this related to reverse engineering models out of the code?}
so that they become explainable, and support decision making.

Glamorous Toolkit (GT)~\cite{feenk23a} is an example of a \emph{moldable} \emph{development} \emph{environment} in which the development tools are continuously molded to make the software under development explainable.
The examples we use to illustrate moldable development in this paper are all drawn from GT.\footnote{GT is open source, and runs on Mac, Linux and Windows. You can download GT from \url{https://gtoolkit.com}.
There you will find alternative descriptions of these patterns in the ``Glamorous Toolkit Book'', a live notebook containing the examples shown here.}

\begin{figure}[h]
  \includegraphics[width=\columnwidth]{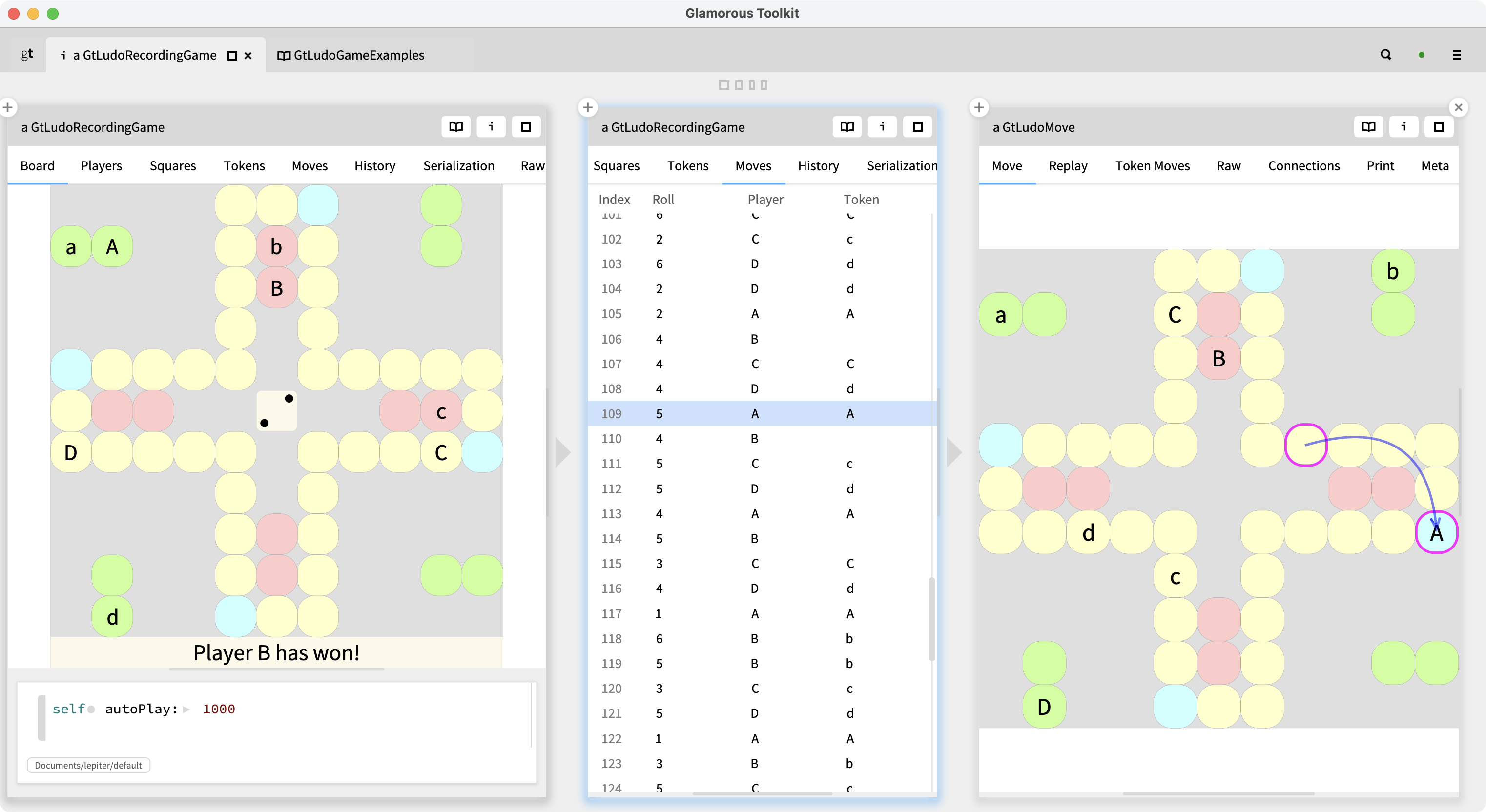}
  \caption{Custom views of a Ludo game.}
  \label{fig:ludoViews}
\end{figure}

As a simple example, consider the inspector views of a Ludo game implementation in \autoref{fig:ludoViews}.
With a conventional implementation, we can either try to play the game interactively, or we can stare at the source code.
We can also run the tests, but if these are all green, they do not help us to understand the system.
By applying moldable development, we turn questions we have about the game into custom views.

The figure shows three connected custom inspector views
%\cp{are these views released to be available to all users of the game? or are they only present during development? -- just trying to reflect on the "your software" or "your software development environment" comment from the beginning of the introduction}
of a running game in GT.
In the leftmost pane we see the game GUI as the \st{Board} view.
We can also interact programmatically with the game, evaluating ``\st{self autoPlay: 1000}'' in a contextual playground (a kind of interactive shell) below the view.
In the second pane we can explore the moves of  the completed game, and in the third pane we can explore individual moves.
The \st{Move} view visualizes the actual move performed in the context of the current game state at that point in time.
Each of these views is achieved with just a few lines of code, and leads to the Ludo game becoming an explainable system that can be explored in ways that are far richer and more intuitive than by trying to read source code.

In a nutshell, moldable development is a way to make systems explainable by making the \emph{inside} of a software system visible and explorable through custom tools.
%\cp{how many of these tools are domain/application specific and how many are reusable across different applications?}
Of course, this begs the question how to actually apply moldable development in practice.
In our experience applying moldable development to many industrial and open source systems, we have encountered a number of repeating patterns, which we document below.

% ===== Moldable Development Patterns =========================
\section{Moldable Development Patterns}

%\rb{In our view, the paper is weakened by the assumption that it is aimed at the same audience as the previously published paper. Your goal in the "book" was to inform GT students of the scope of possibilities in this product. We would hope that your goal in publishing samples through PLoP would be to spread your success more broadly. (Ward is thinking of sharing with him for example). }

%\dd{I don't necessarily agree with the pattern collections being pattern languages --- the relationships between patterns also matter.}

\begin{figure}[h]
  \includegraphics[width=\columnwidth]{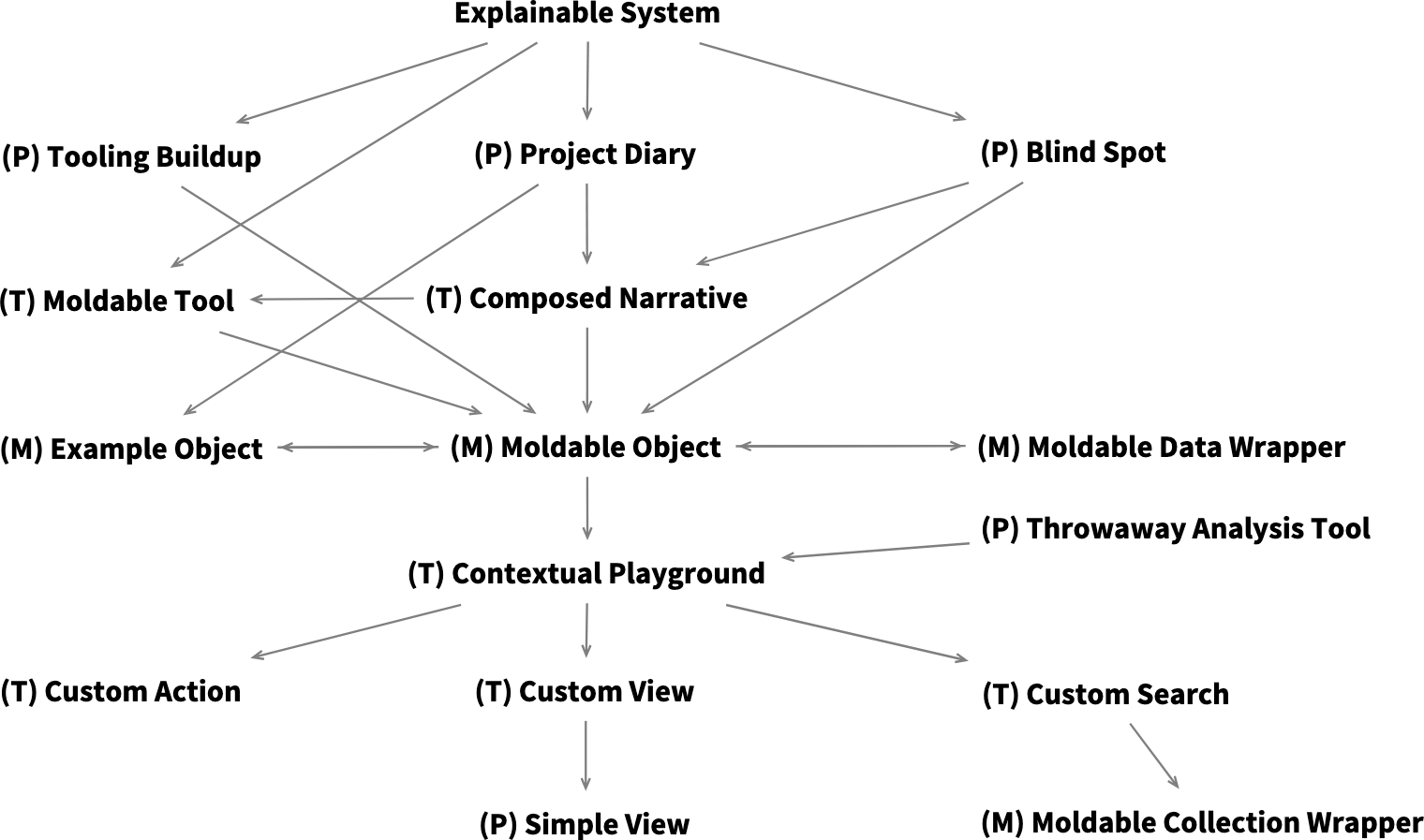}
  \caption{A map of moldable development patterns.
%  \cp{consider adding a legend for the arrow. Having "it depends on" relationships helps really well to build up the hierarchy until the goal of having an explainable system. The question is whether explainable system requires ALL or AT LEAST ONE of the underlying patterns, likewise for Contextual Playground - does it need ALL or AT LEAST ONE of the custom view/search/action patterns.}
  }
  \label{fig:map}
\end{figure}

Moldable development can be understood in terms of a collection of mutually supporting 
%\dd{Make this more evident}
patterns, in other words, a \emph{pattern language}, summarized in \autoref{fig:map}.
These patterns have emerged over several years of experience in developing \GT following moldable development, and applying it to numerous projects.

%\kh{What is not obvious from the description of the patterns is how they mutually support each other. There's an opportunity for a follow-up paper that describes how the patterns are applied in a real project.}

%\ws{JN: The path through the patterns is a crazy maze (shows green lines drawn on the map). (Need to show an idealized path?)\\
%PP: Instead of showing the patterns in the order by the categories, instead just show an idealized order?\\
%JN: Or, like in the GoF book, just list them in alphabetical order.}

%\ws{PP: Must we use all the patterns or is there a minimal subset?
%Why isn't a debugger enough? 
%OK, I see. You can get at the data with a debugger, but it may be hard. The custom views make it easy to get at the answers.}

%\eog{I think like that, the hard thing for this section number two that I found is that there's like a lot of definitions of what things are without a lot of connected motivation.
%And it felt like I don't know if this is a thing that you've ever experienced, but you sit down you play, you're ready to play a board game with some friends and then someone's like, all right, we're gonna sit down and read the rule book from like cover to cover and then everything will be good.
%And then you've got people like me who are like, what is the point of this game?
%How do I win?
%And then like from that I can gather, like my, I can interest my brain in the parts that matter.
%And it's really tough to hold on to just like a list of definitions, for example, without that.}

The diagram illustrates some of the ways in which the patterns support each other:
the arrows indicate that one pattern ``uses'' or ``leads to'' another.
The patterns that are concerned with \emph{Tooling} are tagged with ``(T)'', with \emph{Modeling} ``(M)'', and with the development \emph{Process} ``(P)''.

There are two distinct roles involved in moldable development:
\begin{inparaenum}[(i)]
\item the \emph{Facilitator} is responsible for the technical realization of custom tools, and
\item the \emph{Stakeholder} is responsible for the domain model and questions about the domain that should be answered by the custom tools.
%\cp{can a tool always answer a question? or is a tool used to answer a question?}
\end{inparaenum}
In a purely technical domain, these two roles can often be played by the same person (\ie a developer), but in general they may be distinct people.

At the top of the map we have ``Explainable System,'' which is not a pattern per se, but rather the goal of moldable development.
%\cp{why not consider to promoting "explainable system" to a pattern?}
%\cp{it could be a composite pattern}
An \empty{explainable system} is a software system whose domain models have been exposed with the help of numerous custom tools~\cite{Nier22a}.
%\cp{interesting, but narrow definition. Is exposing its domain model the only way to explain a system?}
%\ws{JN: I would just dump the Explainable System part, and just focus on the moldability. It's also not about reshaping a system.}

Some domains require a preliminary phase of \patref{ToolingBuildup}, for example, to create dedicated parsers for programming languages, DSLs or specialized data formats, or bridges to other execution platforms.
Each custom tool can be seen as an extension of an existing \patref{MoldableTool}, which can be inexpensively customized.
A \patref{ProjectDiary} is a notebook that serves as a starting point for development tasks and recording explanations.
An explanation generally takes the form of a \patref{ComposedNarrative}, a story built up with the help of  live objects.
A \patref{BlindSpot} is a problematic part of the target system that is hard to understand and work with, and may be a promising starting point for moldable development to initially engage Stakeholders.
A \patref{ThrowawayAnalysisTool} can be a quick way to solve an urgent problem in a focused way.
%\cp{What about more heavy-weight tools which perform static/dynamic analysis?}

At the center of the diagram we see \patref{MoldableObject}, which is also the most central pattern in terms of methodology.
Moldable development itself starts with a moldable object, a live instance of a domain entity that is explored and molded with custom tools that package the results of exploration tasks.
%\cp{how is this related to a blackbox, glassbox, whitebox approach? how invasive is moldable development? does it change the system being explained or just helps to observe it? or explanation inevitably leads to evolution...}
An interesting instance can be encapsulated as an \patref{ExampleObject}, essentially a unit test that returns a tested object.
An example can be embedded in a project diary notebook page, and can also be used as a moldable object itself for further development tasks.
In case the domain includes already existing data entities, each of these can be wrapped in a \patref{MoldableDataWrapper} to produce a moldable object.

A moldable object can be explored with the help of its \patref{ContextualPlayground}, a live programming environment bound to the state of a live instance.
Working code can be extracted from such a playground to create custom tools.
The most common of these tools are:
\begin{inparaenum}[(i)]
\item a \patref{CustomView}, a dedicated view of an object within a moldable tool such as an object inspector or a code browser, to display or visualize domain-specific information,
\item a \patref{CustomAction} that encapsulates a useful domain action, and
\item a \patref{CustomSearch}, to perform an ad hoc query over objects reachable from a given moldable object.
\end{inparaenum}
A custom view is frequently a \patref{SimpleView} that can be quickly prototyped, and later extended.
A custom search often benefits from a \patref{CollectionWrapper}, to allow the results of a query to be also molded with custom tools.

%\dd{Good explanation/overview on the pattern language!}

\subsubsection*{How to read the patterns.}
The patterns below are listed in three sections, starting with \nameref{sec:tooling}, and proceeding through \nameref{sec:modeling} and \nameref{sec:process}.
This order reflects the diagram in \autoref{fig:map}, and provides a complete picture of all the patterns.
The hasty reader might prefer to start directly with \patref{MoldableObject} at center of the diagram, and from there follow the arrows to the other patterns, the most fundamental being 
\patref{ExampleObject}, 
\patref{ContextualPlayground}, and
\patref{CustomView}.
This will provide a quick introduction to the most central and essential of the moldable development patterns.

\section{Tooling Patterns}\label{sec:tooling}

We start by presenting the patterns that enable the cheap creation of custom tools.
\pattern{Moldable Tool}{MoldableTool}

%\kh{My main criticism of the overall approach of presenting patterns is that some of them aren't really patterns. For me, a pattern is something that has been repeatedly observed in practice. The very first pattern you present, "Moldable Tool", probably isn't a pattern. How many moldable tools have been developed? I know of one.}

\patsec{Context}
You are developing a software system, and find that the existing development tools fall short in supporting
%\cp{helping you to answer...,  empowering you along your quest to answer...}
domain-specific questions about the software.

\patsec{Problem}
\emph{How can you cheaply and effectively extend the development environment with custom tools that address questions about your application domain?}
%\cp{domain-specific is pleonastic with the last part of the sentence}

\patsec{Forces}
\begin{itemize}[---]
\item Generic software tools are fine for answering generic questions, but they do not work well when addressing domain-specific questions.
%For example, consider a generic debugger being used to debug an event-driven application\,---\,you want to step through the chain of events, not the stack.
%\dd{Consider separating the forces, eg with a line break.}
\item A plugin architecture can open up an IDE to new tools, but plugins can be complex and expensive to implement, 
%\cp{while cost is a clear force, "play nicely" could be more sharply defined. Is it a compatibility issue? Eclipse/VS code are complex, but have well designed extension points with a clear separation of tool boundaries and interfaces. Is coupling between a "moldable tool" and the software on which it is applied a concern? if the internals get exposed through the tool, doesn't this violate information hiding?}
and they often do not compose nicely with existing tools or with each other.
\end{itemize}

%\tk{We call it "the preplanning problem" when discussing frameworks + plug-ins as an implementation option for software product lines. You need to identify all the hot spots for potential extensions in advance, define appropriate extension points, define interfaces which are not too broad and not too narrow, etc.}

\patsec{Solution}
%\cp{Consider splitting the solution from the example/known use. An example is included in most patterns, so the template of the pattern could feature "Example" between "Solution" and "Consequences".}
%\ws{PP: Problems are short and succinct but the solutions not. Perhaps split into Solution and Details.}
\emph{Make the development tools \emph{moldable} to the \emph{dynamic context} of the artifacts they are intended to work with~\cite{Chis17a}, by associating custom behavior to the artifacts themselves.}
%\ws{JN: P. 7, line 119. Since “moldable” and “dynamic” are in italics. Are they forces? There is a Known Use within the solution.}

\patsec{Examples}
A moldable tool makes its core functionality configurable by means of lightweight mechanisms.
For example, a Test Runner in a modern IDE recognizes the presence of test cases by means of various programming conventions, \ie a test case is a method with a standard annotation, or it's a method whose name starts with ``\st{test}'' and belongs to a class that inherits from a \st{TestCase} class or implements a \st{Test} interface.

\begin{figure}[h]
  \includegraphics[width=\columnwidth]{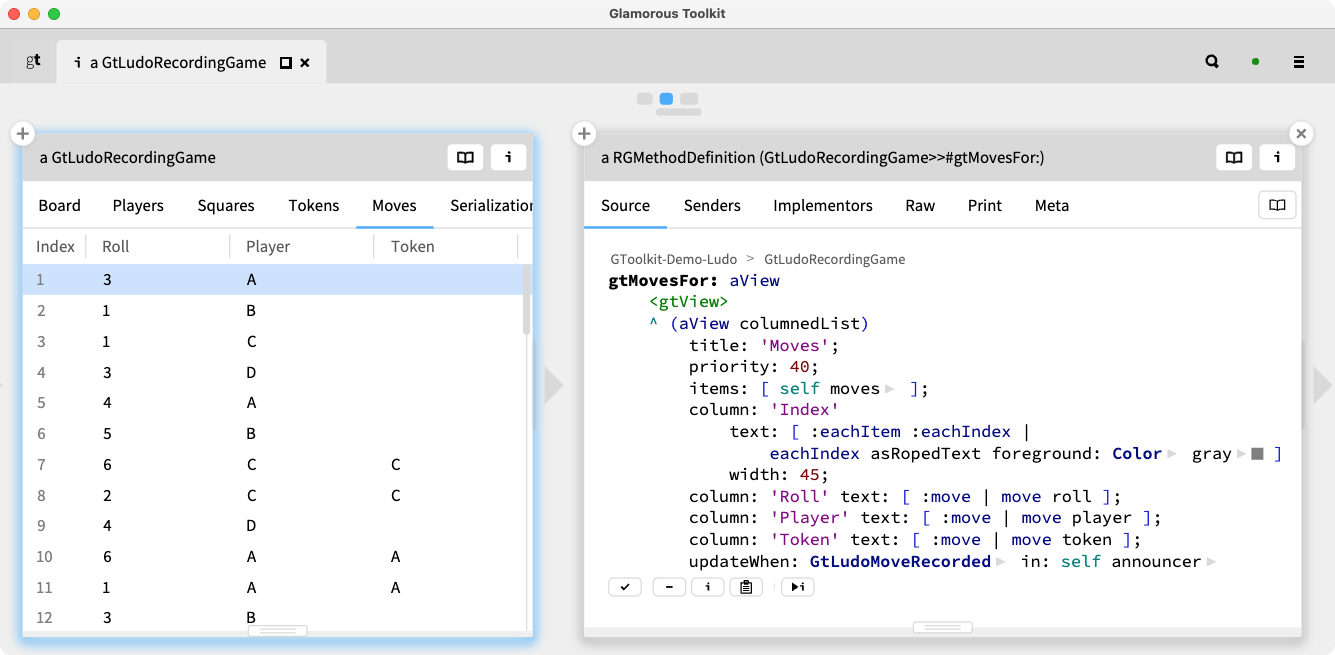}
  \caption{Defining a custom view in a just a few lines of code.}
  \label{fig:viewCode}
\end{figure}

By the same token an inspector can be made moldable by recognizing that an object it is inspecting has one or more custom views defined as annotated methods.
%\cp{so the object has to play nicely with the inspector tool}
For example, in \autoref{fig:viewCode} we see that inspecting an instance of a \st{GtLudoRecordingGame} yields a custom \st{Moves} view listing the moves played thus far, as the inspector detects a \st{gtMovesFor:} method defined in the object with a \st{<gtView>} pragma (\ie annotation).\footnote{All the examples are written in Pharo Smalltalk, the platform upon which \GT is built. See \url{https://pharo.org}}

The precise mechanism used to mold a tool is not as important as is the fact that the customizations should be cheap, \ie few lines of code, and \emph{dynamic}, \ie detected at run time.
%\cp{this powerful "dynamic" aspect could be expanded: the m. tool can expose (or modify) the dynamic runtime state; the m. tool activation does not require to restart/reset the software being explained...}

Other examples include:
\begin{inparaenum}[(i)]
\item a moldable code browser offering alternative views of packages, classes or methods to show dependencies, tests, or other more domain-specific features, such as custom visualizations,
%\cp{such as?}
\item a moldable debugger to adapt the stepping behavior for diverse domains such as event-driven applications or parsing rules, 
% \cp{is this an not numbered standalone example?}
or
\item a moldable notebook, supporting domain-specific snippets or annotations.
\end{inparaenum}

\patsec{Consequences}
A moldable tool requires no up-front configuration, since it will be dynamically molded by the artifacts it encounters.\\
Conversely, since custom tools are an intrinsic part of software systems rather than the IDE, molding happens when needed.\\
Some tooling effort is required to open up the existing IDE tools to make them moldable, \ie dynamically customizable.

%\ws{PP: Consequences mostly seem to be missing negative ones.}

\patsec{Known Uses}
Unit testing tools such as JUnit~\cite{Beck98a} rely either on inheritance or annotations to automatically detect which methods in a package are tests, and adapt the IDE to enable test runners, however the tool adaptation is fixed, and cannot be further customized without writing a new plugin.
The JetBrains IDE allows you to customize the views of objects within a debugging session, though the custom views are specified in the configuration of the tool itself, not in the context of the object.\footnote{\href{https://web.archive.org/web/20240523010248/https://www.jetbrains.com/help/idea/customizing-views.html\#renderers}{https://www.jetbrains.com/help/idea/customizing-views.html\#renderers}}
The Glamorous Toolkit includes several moldable tools.
The Moldable Inspector~\cite{Chis14a} installs custom views, actions (\ie buttons), and search interfaces whenever an object is inspected that defines such custom tools as annotated methods.
The Moldable Coder~\cite{Syre17a} (\ie code editor) also can add custom views, highlighters, actions and searches when editing the source of a class with annotated methods.
The Moldable Debugger~\cite{Chis24a} provides custom views, actions and searches installed as methods in the class of a caught exception.

%\ws{JN: Known uses are needed. If the paper was about explainable systems, a known use could be: A traveler arrives at Heathrow and is denied boarding. The explainable system shows why this has happened.
%If it's the Ludo game, it could be Johnny who loses, and the explainable system explains to him why.
%If it's an industrial system, say that “In System blah we were able to …”.}

\patsec{Related patterns}
A Moldable Tool reacts to a \patref{MoldableObject}, which triggers custom tools such as \patref{CustomView}, \patref{CustomSearch} or \patref{CustomAction}.
A Moldable Tool should provide a \patref{ContextualPlayground} to enable a live programming interface to the entities it works with (\ie objects, classes, exceptions).

%: ----- Contextual Playground -------------------------
\pattern{Contextual Playground}{ContextualPlayground}

%\tg{I would rename this to Contextual Snippet. It's not the page, but the snippet that is essential. We should also say that it is interesting to have different kinds of snippets, much like we can have different kinds of views.}
%\kh{"Contextual Playground" is a pattern, but I'd label it "process" rather than "tooling". On the other hand, it's also reasonable to have it right before "Custom View", as it's a preparation to tooling work.}

%\todo{Not so clear what problem is being solved. Is this about exploring, coding, testing or prototyping? It's all mixed up. What is the insight?}

\patsec{Context}
You are using a \patref{MoldableTool}, and you are ready to start exploring your software system.

% \eog{So you're exploring maybe we could talk about what, what exploring means. Like if you ask a regular developer, like tell me about the explorations you've done this morning, versus, hey, what kind of like debugging or like learning about the system?}

\patsec{Problem}
\emph{Where do you start exploring?}
%\cp{is coding a synonym of exploring? or what is the relationship between the two?}

\patsec{Forces}
\begin{itemize}[---]
\item An editor for coding new methods typically provides no facilities for testing the code.
\item When we code new behavior as methods, we must repeatedly change our context to incrementally develop the logic.
\item Testing the code requires a separate setup.
%\cp{This text is very concrete, trying to give an example where a fragmented tool UI design can lead to poor usability.}
\item Setting up code to prototype and test logic can be cumbersome.
\item Writing tests first for parts of the logic of a complex method can be overkill.
%\cp{isn't this what test-driven development advocates?}
\end{itemize}

\patsec{Solution}
\emph{Use a contextual playground of the tool to programmatically explore the entities of interest, and prototype new behavior.}

%\cp{this pattern has not been introduced yet, so readers may need to jump ahead to learn about it before understanding the solution of the contextual playground}

\patsec{Steps}
A contextual playground provides a live programming interface to programmatically interact with the software entity managed by a \patref{MoldableTool}.
%\del{Typically this entity will be a live object, but it could be a class, or any other kind of software entity.}
The playground will be bound to the context of the object, so \st{self} and all slots (\ie instance variables) can be accessed exactly as they would in a running method.

Use the contextual playground to write code snippets to extract data, test hypotheses, navigate to parts or other related objects, to explore the API, or to prototype new behavior.
Evaluating the code will typically open a new instance of another moldable tool on the result returned by the code, with a new contextual playground.
Code snippets that turn out to be useful or interesting can then be copy-pasted to existing methods, or extracted to new methods using an \emph{Extract method} refactoring~\cite{Fowl99a}.
The same process can be used whether the extracted code consists of accessors, assertions, examples, or new behavior.

\begin{figure}[h]
  \includegraphics[width=\columnwidth]{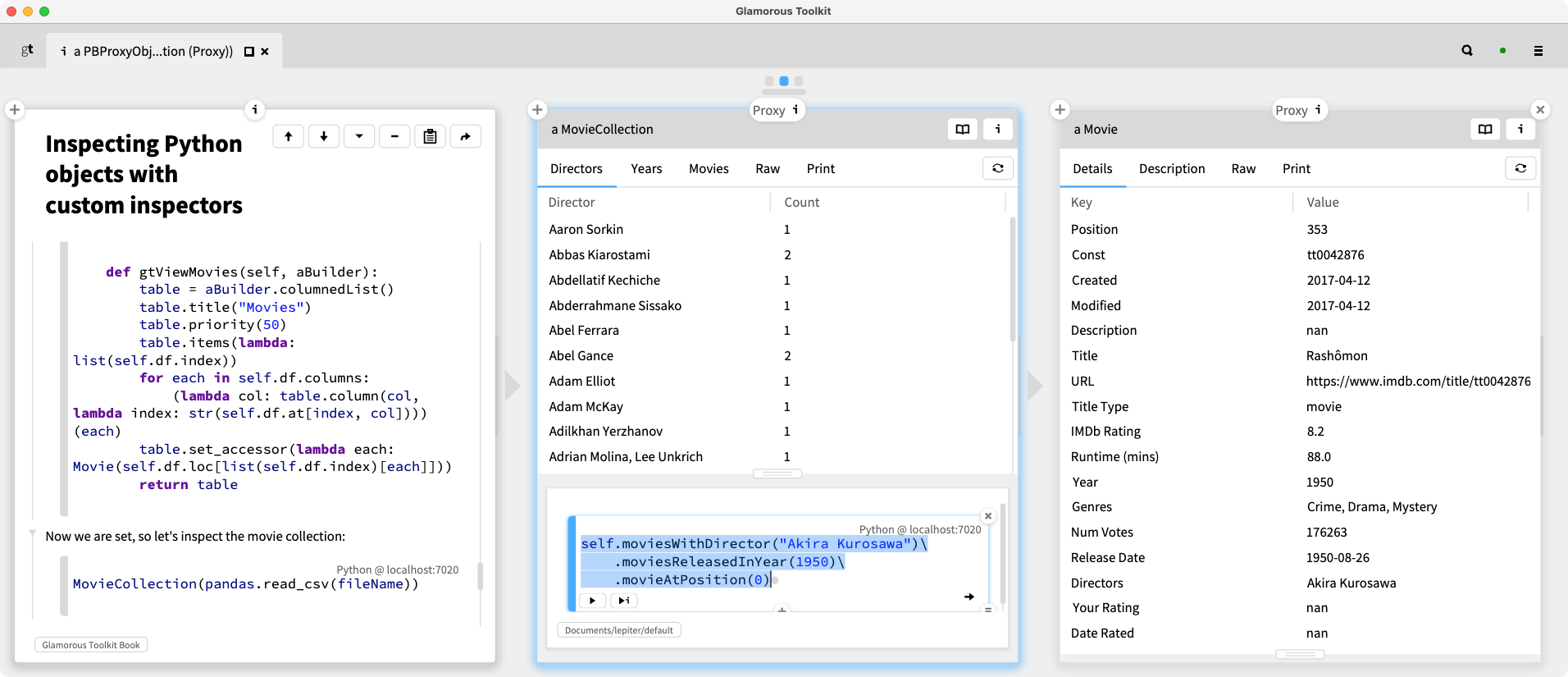}
  \caption{Exploring Python objects within a contextual playground (bottom of middle pane).}
  \label{fig:inspectingPythonObjects}
\end{figure}

\patsec{Examples}
In \autoref{fig:inspectingPythonObjects} we see an inspector view 
%\cp{is this how the contextual playground looks like?}
of an instance of a Python \st{MovieCollection} class.
In the leftmost pane we see a notebook page with live code snippets, which function as contextual playgrounds within the context of the notebook page.
% \tg{I am not sure we should call the global pages contextual. I somehow would keep that word only for the ones in a non-global context - like an inspector or debugger}.
From there we navigate to a moldable object inspector, a moldable tool for exploring live instances.
In this case we are exploring a live Python instance of a \st{MovieCollection} object, with several custom views showing the \st{Directors}, \st{Years} and \st{Movies} of the collection.
At the bottom of the inspector we see a contextual playground bound to the context of the Python object.

%\cp{This section reads more like a "Known Uses" section as opposed to a solution}
In this contextual playground, we experiment with Python code to perform a query over the collection and navigate to a particular \st{Movie} instance.
Note that \st{self} in the code snippet is bound to the context of the \st{MovieCollection} instance.
Once we have identified some useful code, we can extract it as a new method of the object.
Similarly, if we identify an interesting instance, we can use the code to define an \patref{ExampleObject}.

Evaluating this snippet opens another Python object inspector on the resulting \st{Movie}, seen in the third pane.
The third pane also has a contextual playground, which can be opened by dragging up the handle at the bottom of the pane.

\patsec{Consequences}
By prototyping new behavior in a contextual playground, you obtain immediate feedback for experimental code.\\
You may need to evaluate multiple snippets in multiple playgrounds to get to  interesting information.

%\dd{Are there any drawbacks/liabilities/less desirable consequences?}
%\todo{You need an object first.\\
%You may need to evaluate multiple snippets in multiple playgrounds to get to  interesting information.}

\patsec{Known Uses}
A REPL (Read-eval-print loop) or language shell is essentially a contextual playground for a programming language or an operating system.\\
The JavaScript console of a modern web browser also functions as a contextual playground: while inspecting a live web page, you can explore and programmatically interact with the live objects in the run-time environment of the page.\\
The moldable tools within GT all provide contextual playgrounds: notebook pages, object inspectors, code editors, and the debugger, in particular.

\patsec{Related patterns}
%\cp{how is this related to writing tests and the issues mentioned in the forces?}
Use a \patref{ProjectDiary} to start live coding.
You can use a contextual playground to prototype custom tools (see \patref{CustomView}, \patref{CustomAction} and \patref{CustomSearch}) for a \patref{MoldableObject} within a moldable inspector.
A snippet within a contextual playground that yields an interesting object can be extracted as an \patref{ExampleObject}.
A contextual playground for an \patref{ExampleObject} can also be used to explore and prototype assertions for that example.

%: ----- Custom View -------------------------
\pattern{Custom View}{CustomView}

%\ws{PP: I see how this works for user interfaces, but not clear how it works for other domains.\\
%ON: Do you mean that you don’t see how moldable development would work for non-GUI applications?\\
%PP: answer – I don’t see how this would work with an application without a user interface.\\
%(Problem: the paper examples are UI heavy. Over lunch discussed how data-heavy applications, like GitHub actions can also be handled the same way.)}

%\ws{JN: Michael Jackson in one of his books says, “the more applicable something is, the less specific help it will provide.” A lot of this is playing in this space.\\
%ON: Do you mean that the patterns or the tools are too generic? The whole idea of MD is that you need to build custom tools using these general patterns.\\
%JN answer: I’m not sure you really want custom views. One issue is the time and effort to build a custom view. You can spend no time and use a raw view, or you can spend a lot of time building a beautiful custom view. The paper does not get into how you build the views. It seems to be by textual code – not sure why. You could have a view builder that explains views. Collection Wrapper wasn’t clear why it’s needed. There are a lot of assumptions about the language you’re using which aren’t explicit.\\
%Perhaps you should cite Self and Lively.}

\patsec{Context}
You are exploring a live domain model and find some interesting information.

\patsec{Problem}
\emph{How do you make it easy to find interesting information?}
% \tg{I am not certain about the *again* part. Most often we want a view to be useful the first time we use it. That we can use it a second time is a bonus.}

\patsec{Forces}
\begin{itemize}[---]
\item To obtain an interesting view of data, you are generally forced to activate a dedicated tool.
\item Navigating to the data you want to reach may entail a sequence of operations, either clicking in views, or evaluating code snippets, to reach the answer you seek.
%\cp{this is only implicitly hinted at in the previous pattern. Consider explaining it already in the playground pattern}
\item The sequence of steps may be cumbersome to follow repeatedly.
\item The default view you obtain may not highlight the interesting bits of information.
\end{itemize}

% \eog{this part here, you could describe it also with like a watch statement in javascript or whatever, like this kind of thing. And then people could be more familiar with that because they, they do that.}

\patsec{Solution}
\emph{Turn interesting data into a custom view.}
Extract the navigation steps into a new custom view for the moldable object you start navigating from.

%\tg{I think we should emphasize the contextual nature of the view. There is a mapping between the view and some context (object in inspector, stack in debugger and we can imagine other mappings) based on which views are presented in the environment. Because of this, we do not seek the view. The view comes to us.}
%\on{The point is that we do not want to have to always go to a tool to do something, but rather than views will enable the ability to get to interesting stuff just by navigating.}

%\eog{I'd love to see maybe a picture like a figure with this.}

\begin{figure}[h]
  \includegraphics[width=\columnwidth]{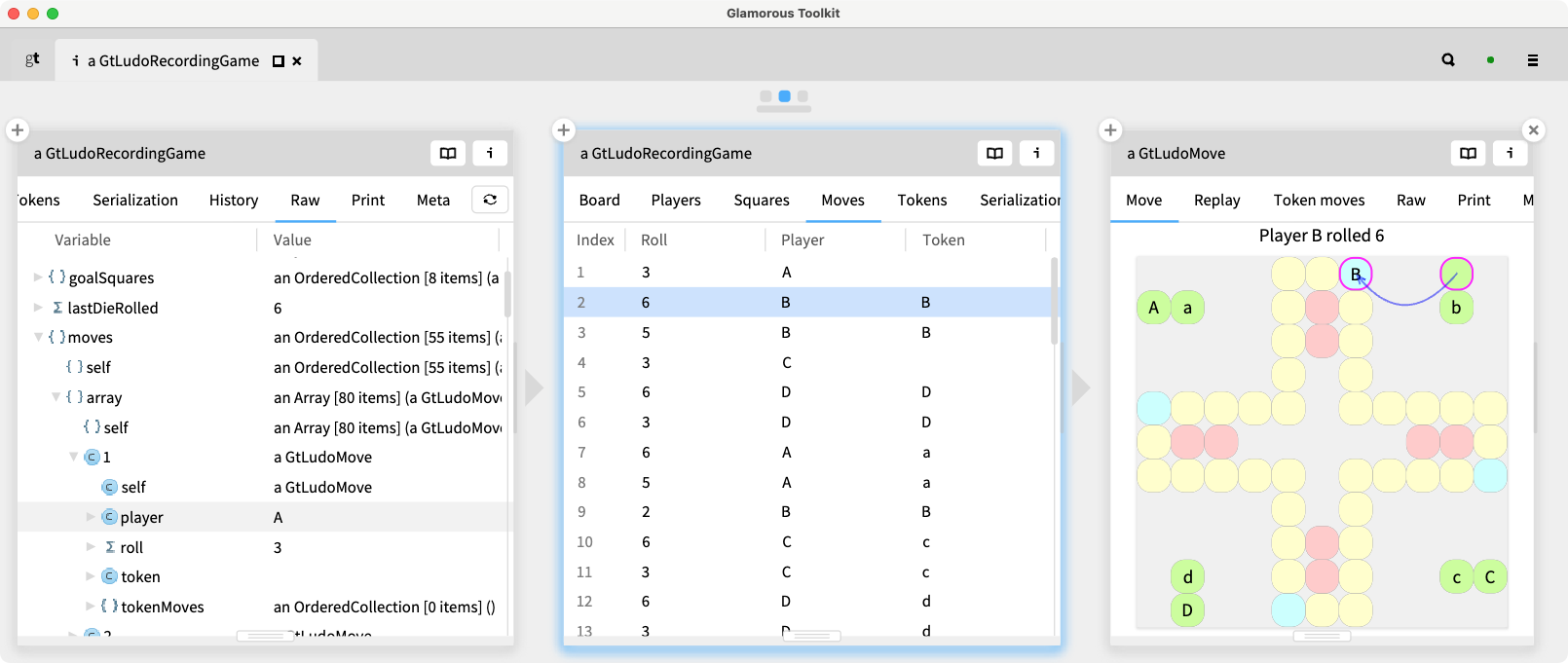}
  \caption{A raw view and two custom views.}
  \label{fig:rawViewVsCustomView}
\end{figure}

\patsec{Examples}
As an example, consider the views of a partially played Ludo game in \autoref{fig:rawViewVsCustomView}.
We would like to understand which moves have been played up to now.
In the leftmost pane we are exploring a ``raw view''\,---\,a generic view of the state of the object.
With this view we can navigate to the individual moves and explore them, but it is a clumsy way to explore the object.
In the second pane we see a custom \st{Moves} view that lists the moves in order, with columns showing each roll of the die, the player who rolled the die, and any token that may have moved.
From this custom view we can then dive directly into the \st{Move} object, which in turn has a custom view visualizing the change in state.

\patsec{Consequences}
A custom view exposes information about a domain that can otherwise be hard to find.\\
Custom views become an intrinsic part of a software system,
%\cp{are those views accessible from end users? -- trying to understand what characterizes them as "intrinsic"}
thus turning it into an \emph{explainable system}.\\
By augmenting an object with a custom view, you can navigate directly to interesting information about that object, without having to fire up a separate tool.\\
You first need a \patref{MoldableTool} into which you can dynamically plug custom views.

%\ws{DP: In the consequences of Custom View there are references to related patterns. pp 10-11. It seems strange to have that in the consequences. Before the consequences mention the related patterns.\\
%PP: Would not put that in the Solution. I would separate related patterns in a subsection in the template.}

%\eog{Down here, there's mention of what a you need a moldable tool, but I still don't understand what a moldable tool is given just this paper so far.}

\begin{figure}[h]
  \includegraphics[width=\columnwidth]{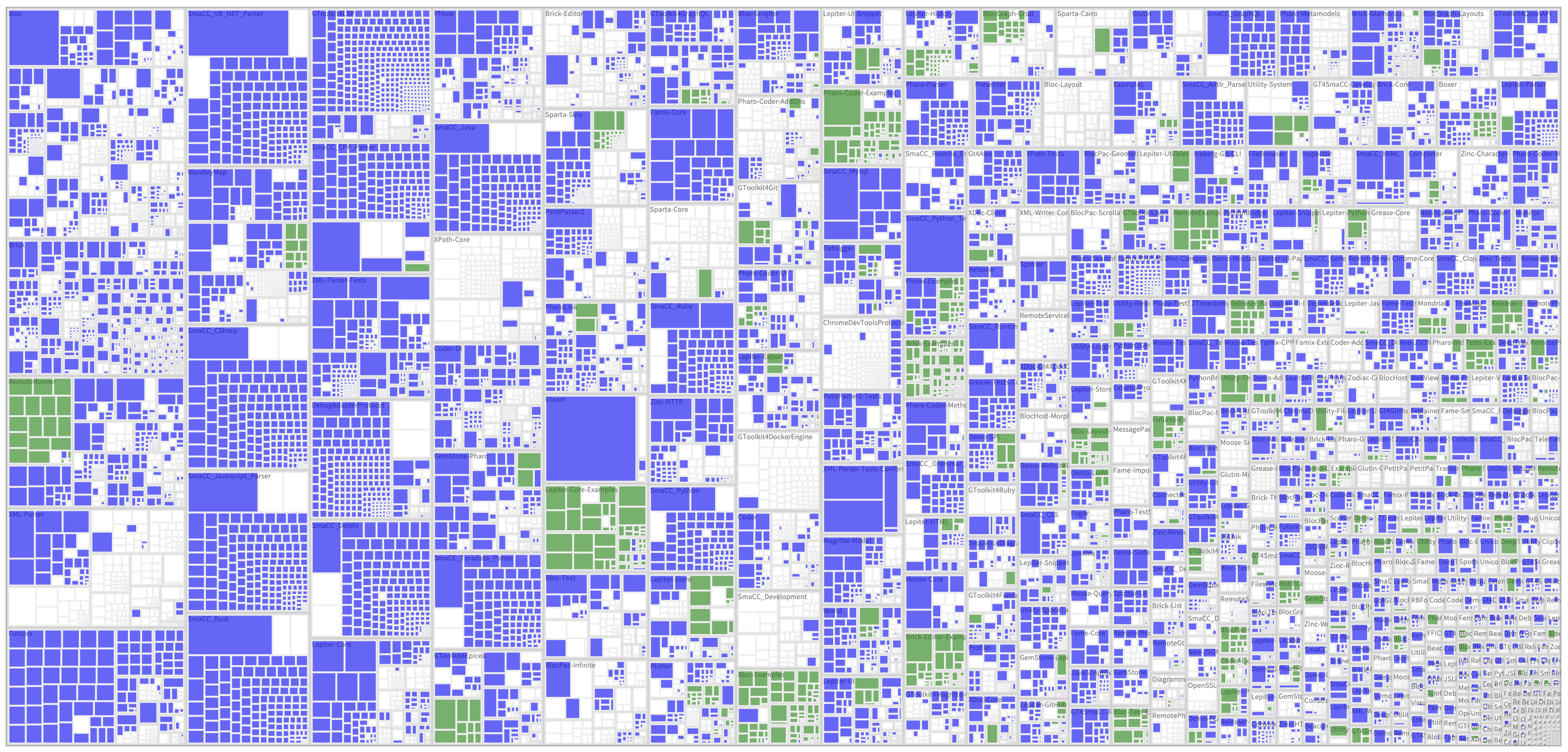}
  \caption{A treemap of all classes in GT grouped in packages, highlighting in blue those that define at least a Custom View.}
  \label{fig:treemapViewsAndExamples}
\end{figure}

\patsec{Known Uses}
Clerk~\cite{Kava23a} is an open-source programming assistant for the Clojure language, which offers moldable, custom views within notebook pages.
Custom views are pervasive in GT.
In a standard image from August 2024, there exist over 1800 classes with a total of over 3600 view methods, averaging under 12 lines of code. 
It is also worthwhile to note that if we take inheritance into account, then the views affect 12000 classes. The treemap from \autoref{fig:treemapViewsAndExamples} offers an overview of the GT classes organized in packages. A class appears is colored with blue if it or its superclass defines a custom view, and in green if it defines an example method (see \patref{ExampleObject}).

% In the image v1.0.1073 of Aug 27, 2024, there are 3627 gtView methods in 1893 classes
%In the image DEV v1.0.1090 2024-09-02T16:07:58.717276+02:00 we also have:
%methods := #gtView gtPragmas result toArray wait.
%(((methods collect: #methodClass as: Set) reject: [ :each | each = Object or: [ each = ProtoObject]]) flatCollect: #withAllSubclasses) reject: #isMeta
%==>12029

\patsec{Related patterns}
A custom view is closely related to a given \patref{MoldableObject}.
Interconnected objects with custom views can form a \patref{ComposedNarrative}.
A \patref{MoldableTool} is a prerequisite for a custom view to be automatically installed when the relevant artifact (an object, a class, an exception, \etc) is seen by the tool.
A \patref{CustomView} should be cheap to implement.
Start with a \patref{SimpleView}, and only elaborate it when it is needed.

%: ----- Custom Search -------------------------
\pattern{Custom Search}{CustomSearch}
\patsec{Context}
%\cp{Why not just send a query to the database where the objects are stored? or are these objects only found in memory at the time they are being searched?}
You have a complex domain model consisting of various kinds of entities related to each other.

\patsec{Problem}
\emph{How can you effectively navigate between domain entities by name, content or other criteria?}

\patsec{Forces}
\begin{itemize}[---]
\item It can be hard to anticipate which domain  entities you will need to navigate to.
\item Designing a good query interface can be a difficult task.
\end{itemize}

\patsec{Solution}
\emph{Add a custom search for every kind of domain entity you want to navigate to.}

\patsec{Steps}
Simply search by substring against various attributes of the domain objects.
Add a separate search for each attribute type or domain object.

\begin{figure}[h]
  \includegraphics[width=\columnwidth]{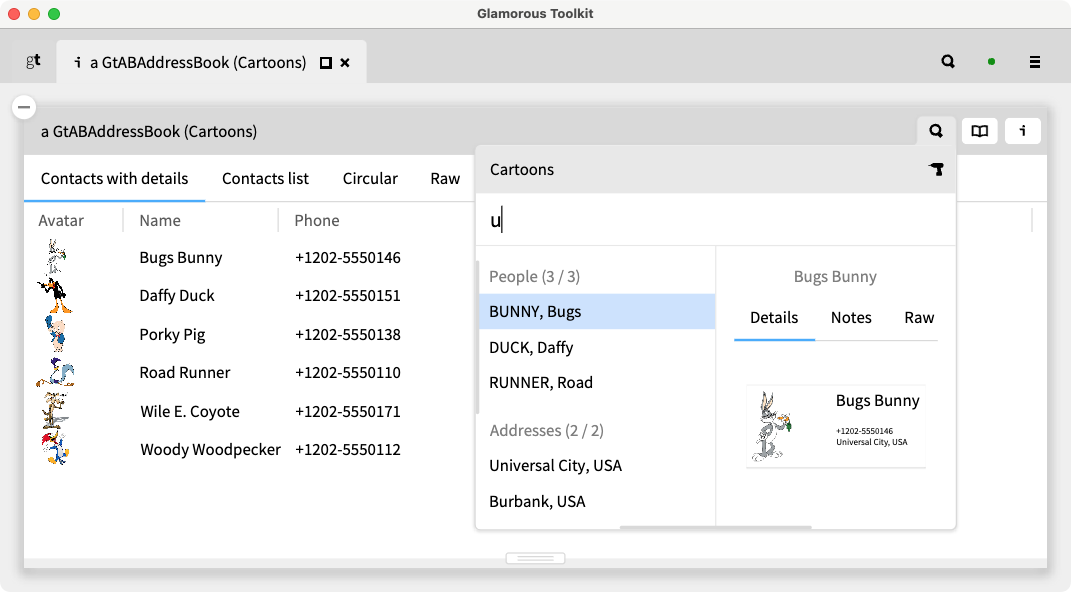}
  \caption{Custom substring searches.}
  \label{fig:addressBookSearch}
\end{figure}

\patsec{Examples}
In \autoref{fig:addressBookSearch} we see custom searches for people and addresses within an address book.
%\eog{I think what I was curious about is if there is a way to use an example of a domain specific custom search example, that would be really cool.
%So I don't know, maybe something in ludo or like something more than just like researching.}
As with other custom tools, one way to implement a custom search is as a method of the domain object from which the search is initiated, with a dedicated annotation.
In \GT, custom searches are annotated with \st{<gtSearch>}, and triggered within multiple moldable tools, such as the inspector, the code editor, and also the notebook.

%\begin{figure}[h]
%  \includegraphics[width=\columnwidth]{CustomSearch}
%  \caption{Custom substring searches. \tg{I would rather pick a more explicit example like the address book. Or at least we should show on the left a collection that we search through.}}
%  \label{fig:CustomSearch}
%\end{figure}

\patsec{Consequences}
You need a \patref{MoldableTool} into which you can dynamically plug custom searches.\\
The same search interface can accommodate multiple custom searches to query different attributes of multiple domain entities.

\patsec{Known Uses}
A form of custom searches are supported by the JetBrains IDE\footnote{\href{https://web.archive.org/web/20240413225707/https://www.jetbrains.com/help/writerside/custom-search-service.html\#request\_format}{https://www.jetbrains.com/help/writerside/custom-search-service.html\#request\_format}} to support configurable help services.
Within GT, each knowledge base of notebook pages supports custom searches by title or contents.
Within the moldable inspector, any object can define multiple custom searches over related collections of objects, for example, a movie database could search for matching titles, directors, or countries of origin.

\patsec{Related patterns}
You need a \patref{MoldableTool} to automatically plug in a custom search.
In order for the result of a search to also be a \patref{MoldableObject}, you should consider wrapping it as a \patref{CollectionWrapper}.

%: ----- Custom Action -------------------------
\pattern{Custom Action}{CustomAction}
\patsec{Context}
You are developing an explorable domain model of your application and find yourself repeatedly evaluating the same code snippets to perform a certain action or navigate to another object.

\patsec{Problem}
\emph{How can you streamline execution of repeated actions?}

\patsec{Forces}
\begin{itemize}[---]
\item Repeated tasks are annoying, time-consuming, and error-prone.
%\cp{error-prone too}
\item Remembering how to perform common tasks increases cognitive overload.
\item Storing code to perform common actions as methods or as snippets in class comments doesn't guarantee that the code will be easily found when you need it.
\end{itemize}

\patsec{Solution}
\emph{Add a custom action button to the moldable tool for the object involved in a repeated task, encapsulating the boilerplate code to perform it.}

\patsec{Steps}
Be sure to pick an evocative button icon and tooltip text to make the intent of the button clear.
Only add buttons for the most important actions to avoid cluttering the interface of a moldable tool.

\begin{figure}[h]
  \includegraphics[width=\columnwidth]{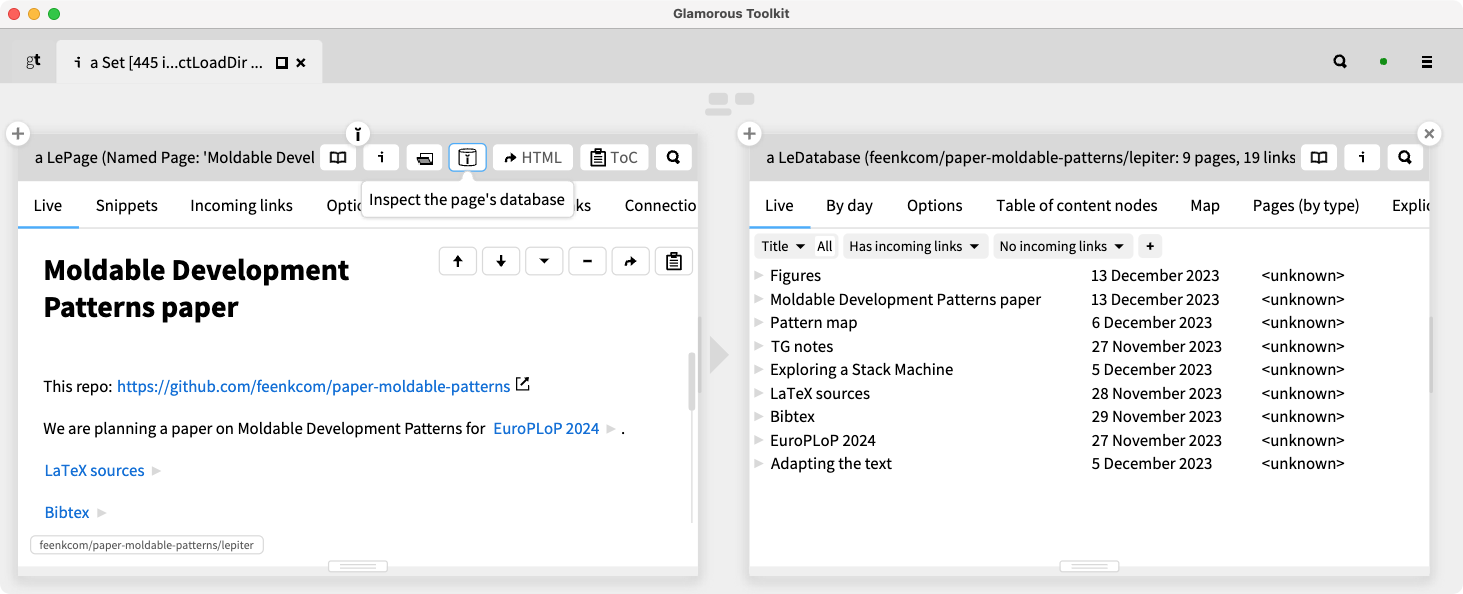}
  \caption{Custom actions for Lepiter pages.}
  \label{fig:customLePageAction}
\end{figure}

\patsec{Examples}
As an example, consider the inspector view in \autoref{fig:customLePageAction} of a notebook page in the GT Book~\cite{Girb21a}, the live documentation system of GT.
Common actions are to navigate to the notebook database, to view the file in which the state of the notebook page is stored, or to export an HTML version of the page.
Each of these actions can easily be packaged as an inspector button, so that the action can be performed with a single click, opening a new inspector view of the result.
In the example we navigate to the database holding all related notebook pages for the given project.

\patsec{Consequences}
Custom actions appear as buttons in moldable tools only in the context of the objects to which they can be applied.\\
You need a \patref{MoldableTool} into which you can dynamically plug custom actions.
%\cp{is this meant to point to related patterns? according to the map in Fig. 2 the Custom View/Action/Search are directly connected with the playground - the moldable tool is further away}

\patsec{Known Uses}
% 314 gtAction methods in 215 classes
In the GT image of August, 2024, there are over 200 classes that define a total of over 300 actions, ranging from checking HTTP links within web pages to pushing commits to a remote repository.

\patsec{Related patterns}
A \patref{CustomAction} can also be used to spawn a particular \patref{CustomView} of the same object or a related one.

%: ----- Composed Narrative -------------------------
\pattern{Composed Narrative}{ComposedNarrative}

\patsec{Context} 
You want to explain a particular aspect of a software system.
It could be an explanation of how to use the system, or a problem or bug that needs to be resolved.

\patsec{Problem}
\emph{How can you explain an issue with a software system in an intuitive way that is easy for a reader to follow?}

\patsec{Forces}
\begin{itemize}[---]
\item Pure textual narratives can be hard to follow and visualize.
\item A picture is worth a thousand words.
\item An individual screenshot without context can be misleading.
\end{itemize}

\patsec{Solution}
\emph{Create a narrative composed of views of various objects or tools, where each view leads to the next by performing a particular action or step.
Share the narrative as a static screenshot, record it as a video, or perform it live with an audience.}

\patsec{Steps}
Choose a starting point within a given tool, such as a notebook page, an inspector view of an object, a debugger window, or a source code editor on a class or package.
Navigate to a new state by performing a step, such as clicking on an element of a view, performing a custom action, initiating a custom search, evaluating a code snippet in a contextual playground, or any other GUI action.
String together these states and steps into a narrative.
Capture the narrative as a screenshot or a video, or a live, scripted GUI entity, and share it.
Be sure to capture the entire view of each tool or object to provide the necessary context.
If necessary, edit or decorate the narrative to highlight certain aspects or steps performed.

\begin{figure}[h]
  \includegraphics[width=\columnwidth]{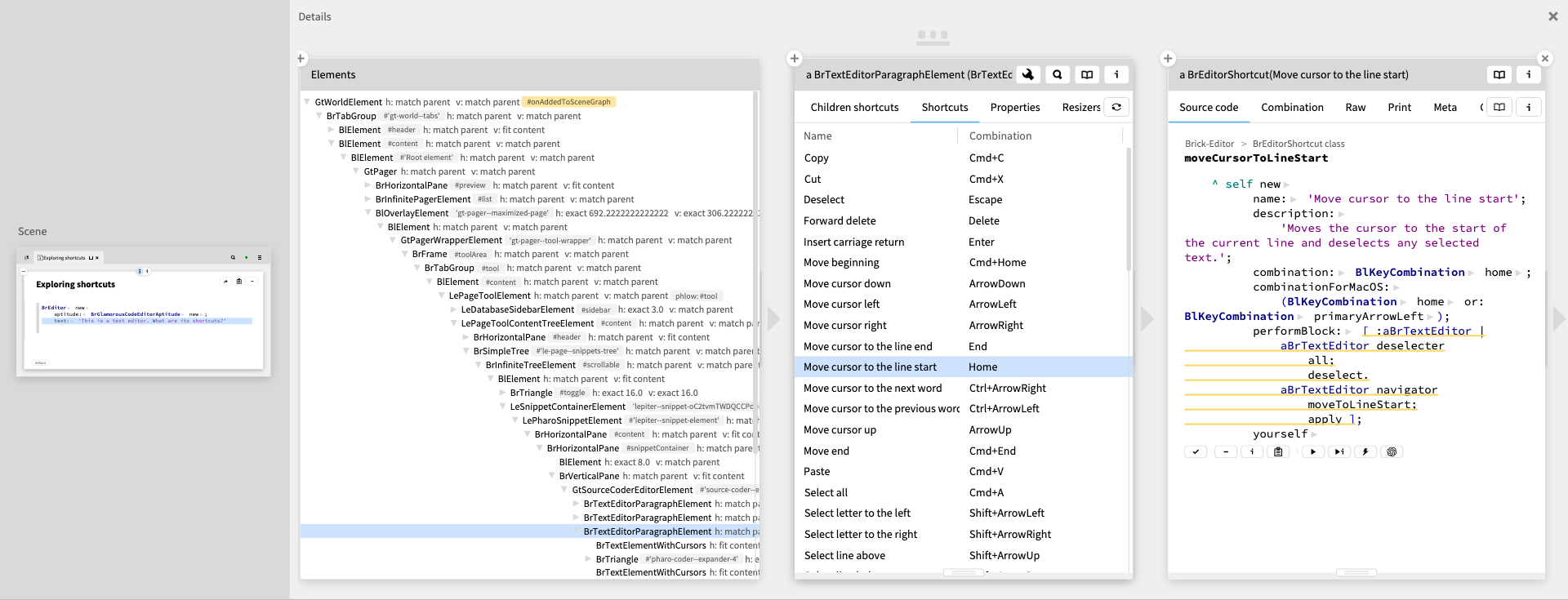}
  \caption{An composed narrative showing how to explore the shortcuts available in an editor.}
  \label{fig:drillerShortcuts}
\end{figure}

\patsec{Examples}
In \autoref{fig:drillerShortcuts} we see a composed narrative answering the question of how to locate the implementation of a shortcut of an editor. The narrative starts with a visual scene for which we see the rendering tree (similar to an HTML DOM). A paragraph object from the editor is selected and reveals its list of shortcuts. A shortcut object presents its defining source code and concludes the answer.

Most of the figures in this paper are illustrations of Composed Narrative.
For example, \autoref{fig:viewCode} shows how a given Custom View is implemented in a few lines of code.
\autoref{fig:rawViewVsCustomView} shows how a raw view differs from a Custom View.
\autoref{fig:composedExample} illustrates how examples can be composed to create new examples.

\patsec{Consequences}
A Composed Narrative tells a story of how one can get to an interesting state.
You may need to carefully adjust the individual views to highlight what is interesting.
Composed Narratives consisting of more than three views may be unwieldy when shared as static screenshots.

\patsec{Known Uses}
Narrative art works,\footnote{\href{https://web.archive.org/web/20240916005104/https://en.wikipedia.org/wiki/Narrative\_art}{https://en.wikipedia.org/wiki/Narrative\_art}} from the Bronze Age through to modern graphic novels, tell stories by juxtaposing images.
Montage Theory, dating from the early days of cinema~\cite{Eise49a}, creates narratives in film by creatively juxtaposing individual shots.

\patsec{Related patterns}
The first view of a composed narrative is often a \patref{ProjectDiary} notebook page, or a \patref{CustomView} of an \patref{ExampleObject}.
The steps performed may be a \patref{CustomAction}, a \patref{CustomSearch} or evaluating a snippet in a \patref{ContextualPlayground}.
A scripted Composed Narrative can itself be turned into an \patref{ExampleObject}.

% ===== Modeling =========================
\section{Modeling Patterns}\label{sec:modeling}

Next we present the patterns related specifically to modeling domain entities in a way that facilitates querying and exploration of an explainable system.

%: ----- Moldable Object -------------------------
\pattern{Moldable Object}{MoldableObject}

%\ws{JN: What's a Moldable Object? (Chants problem solution.) Solution is very heavy - there's a lot of stuff going on here.\\
%DP: The issue with the solution is that it seems to be self-defined. (A Moldable Object starts with a Moldable Object.) We need to take action from the beginning.}

\patsec{Context}
You are ready to start the process of creating an explainable system, either from scratch, or based on some existing software or data, to answer specific questions you have.
%\cp{should a question to be explained already be there? or will the question/answer emerge by itself after this process starts?}

\patsec{Problem}
\emph{Where do you start coding an explainable system?}

\patsec{Forces}
\begin{itemize}[---]
\item As a programmer, you want to quickly get feedback about the code you are writing.
%\cp{get feedback from whom? (maybe) understand the consequences/impact of the code you are writing.}
\item When you write code in a conventional code editor, you are several steps away from seeing the consequences of your coding.
\item To write unit tests, you must already know what behavior you want to test and what the results should be.
\end{itemize}

\patsec{Solution}
\emph{Start coding by inspecting a live instance of the class you are coding, not in a conventional source code editor}.
%\cp{is any live instance of a class already a moldable object?}

\patsec{Steps}
Incrementally pose domain-specific questions, find the answers by exploring and interacting with the object, and then turn the way you reach those answers into custom tools, behaviors and tests.
%\cp{the way to obtain those answers (?)}

\patsec{Examples}
Moldable development is about making systems explainable with the help of custom tools, which means that you need to start the process by asking questions that you want to answer.
In most cases you can't immediately start building the custom tools, but rather you need to explore the domain objects to understand how to answer the question.
Once you know how to get the answer, you can turn the exploration steps into a custom tool.
Starting with a live object means that you can immediately start the exploration process.
Turning the exploration of an object into a custom tool is the process of molding it, hence we call it a ``moldable object.''

\begin{figure}[h]
  \includegraphics[width=\columnwidth]{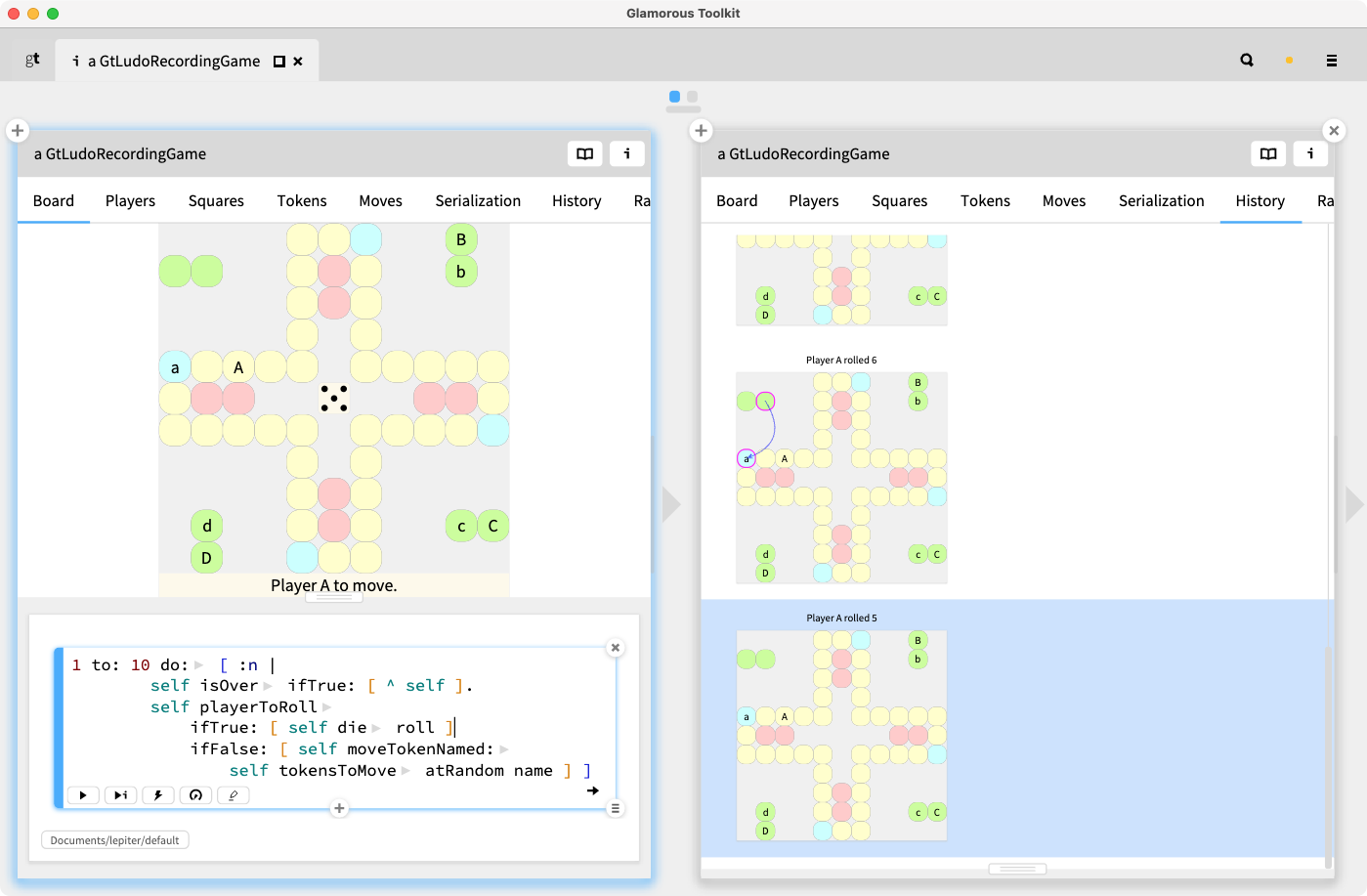}
  \caption{A moldable Ludo game instance.}
  \label{fig:moldableLudoGame}
\end{figure}

In \autoref{fig:moldableLudoGame} we see an Inspector on an instance of a \lst{Gt\-Ludo\-Recording\-Game} with a \patref{ContextualPlayground} where we can write experimental code that we later extract as methods.
In this case we are prototyping the \emph{autoplay} feature.
In the second pane we see the \emph{History} view of the object, showing the details of all the moves of the game.

%\ws{JN: In the Ludo game, where did the history come from? Is it the game or the molding?}

With the moldable object at hand, we can ask questions like \emph{What is the current state of the Ludo game?}, or \emph{What happened in the last few moves that we autoplayed?}
As we answer questions we can create small custom tools, such as a \patref{CustomView} or a \patref{CustomAction}.
The model of the history emerged from the need to answer questions about the evolution of an instance of the game, while the history view was created as a \patref{CustomView} to enable exploration of a game's history.
Whenever we identify an interesting state of our moldable object, we can extract it as an \patref{ExampleObject} that we can use as a test case, or as a starting point for further moldable development.

\begin{figure}[h]
  \includegraphics[width=\columnwidth]{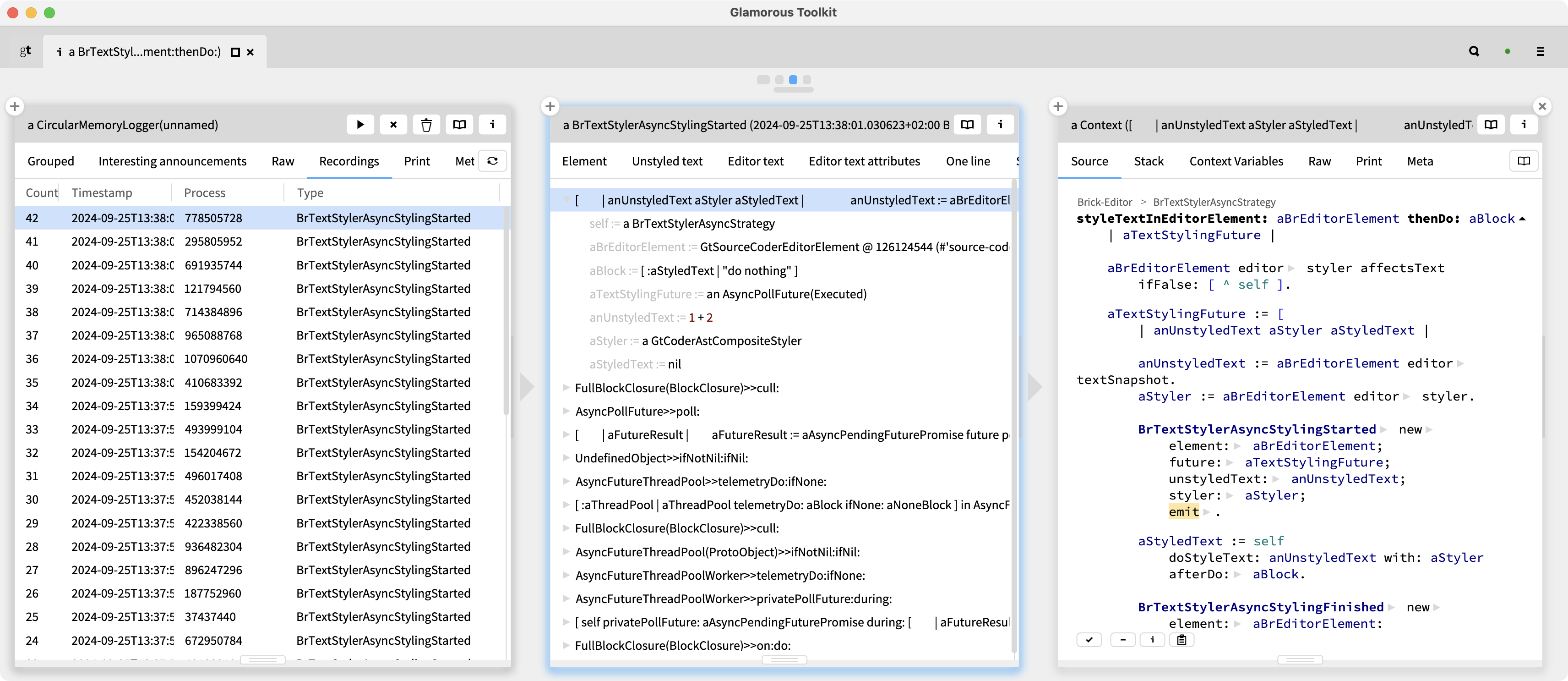}
  \caption{Three interconnected moldable objects playing the role of a logger linked to a debugger.}
  \label{fig:moldableStackCode}
\end{figure}

A moldable object can encapsulate entities at different levels of abstractions.
When viewed with a \patref{CustomView} and connected with other moldable objects we can form various kinds of \patref{ComposedNarrative}.
For example, in \autoref{fig:moldableStackCode} we see three related objects.
On the left we see a \lst{Circular\-Memory\-Logger} object that shows multiple signals logged for an asynchronous execution.
In the middle we have an instance of \lst{Br\-Text\-Styler\-Async\-Styling\-Started}.
The signal contains the stack and presents it in a \patref{CustomView}.
Selecting an item in the stack shows the related \lst{Context} object that displays a view with the Source.
Essentially, we have obtained a postmortem debugger connected to a logger.
There are many such combinations possible.

\patsec{Consequences}
Instead of writing code in a text editor in the context of the source code of a class, you are always working in the context of a live object, whose behavior can be immediately explored.\\
Instead of writing hypothetical code that you must afterward test, you start by prototyping code, and then extracting new behavior.\\
Instead of trying to program custom tools in a vacuum, you first explore and prototype answers to questions, and then extract the code you need to create a custom tool.

\patsec{Known Uses}
The core idea of a Moldable Object is to incrementally mold a live object to obtain immediate feedback into code changes.
Interactive, or ``live'' programming has a long history~\cite{Rein18a}.
One could argue that any live programming task starts with a ``moldable object'', however the key difference is the focus on molding objects to create an explainable system consisting of custom tools, rather than just live programming in general.

\patsec{Related patterns}
You can obtain a moldable object in various ways.
\begin{itemize}[---]
\item \emph{You already have a class:} create a \patref{ProjectDiary} notebook page containing a code snippet to create an instance of the class, and start from there.
\item \emph{You don't have a class:} within a \patref{ProjectDiary} page, start instead with a code snippet that instantiates an empty class, and then prototype the behavior.
\item \emph{You have a test case:} turn the test case into an \patref{ExampleObject} and start from there.
\item \emph{You have some data (in memory, in a file, in a database, in the cloud ...):} 
%\cp{is the data in a file on disk? or is it some data value of a variable in memory? trying to understand what "data" means vs. objects and classes mentioned before}
wrap the data as a \patref{MoldableDataWrapper}.
\end{itemize}

%: ----- Example Object -------------------------
\pattern{Example Object}{ExampleObject}
\patsec{Context}
You want to explore questions about domain objects that are in particular execution states.

\patsec{Problem}
\emph{How do you create an object in a particular state to start a moldable development task?}

\patsec{Forces}
\begin{itemize}[---]
\item Concrete examples are needed for many purposes, such as documentation, testing, and exploration.
\item Examples can be complex to set up.
\item Unit tests consume examples, but they are only accessible if a test fails.
\end{itemize}

\patsec{Solution}
\emph{Wrap examples as (instance) methods that optionally evaluate some tests (assertions), 
%\cp{how do these assertions help answering the problem question above?}
and return the example instance.}
%\cp{sounds like a GOF factory method?}

\patsec{Steps}
Each example may also use one or more existing examples as the initial setup to arrive at the new execution state.
To start, you need a modified unit testing framework in which tests return the exercised fixture, namely, an example.

\patsec{Examples}
In GT, you create an example by defining a parameterless method that has a \st{<gtExample>} pragma and returns an object.
Here is a simple example method that creates a fresh instance of the \st{gtLudoGame} class, asserts a few basic facts (\ie that the game is not yet over, no one has won yet, and so on).
It resembles a classical unit test in all respects except one: it returns the instance of the unit under test, \ie the \st{game} instance.

%\dd{monospace font?}

\needlines{10}
\begin{code}
GtLudoGameExamples>>#emptyGame
	<gtExample>
	| game |
	game := self gameClass new.
	self assert: game isOver not.
	self assert: game winner equals: 'No one'.
	self assert: game currentPlayer name equals: 'A'.
	self assert: game playerToRoll.
	self assert: game playerToMove not.
	^ game  "return the game instance"
\end{code}

\begin{figure}[h]
  \includegraphics[width=\columnwidth]{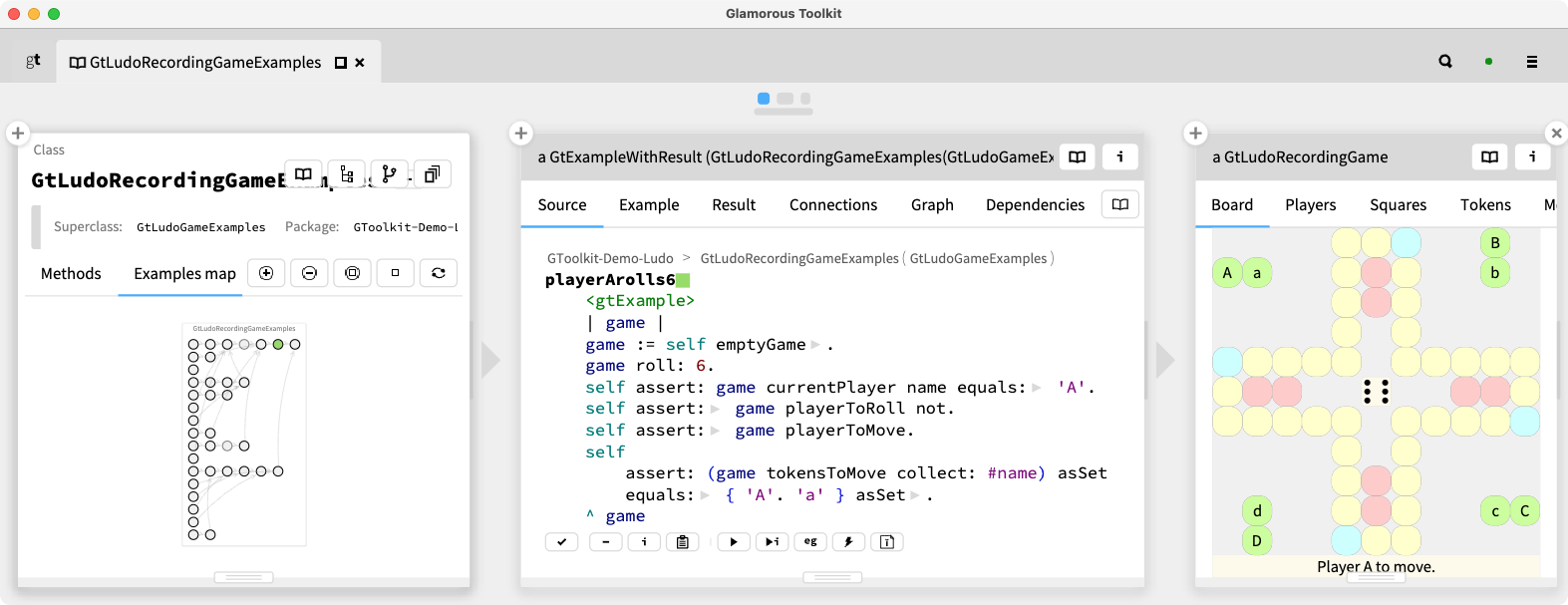}
  \caption{Composing examples.}
  \label{fig:composedExample}
\end{figure}

Unlike normal test methods, examples are designed to be \emph{composed}.
In \autoref{fig:composedExample} we see a second example, \st{playerArolls6}, that starts from the \st{emptyGame} example, rolls a $6$, asserts a few facts, and returns the modified game instance.

\begin{figure}[h]
  \includegraphics[width=\columnwidth]{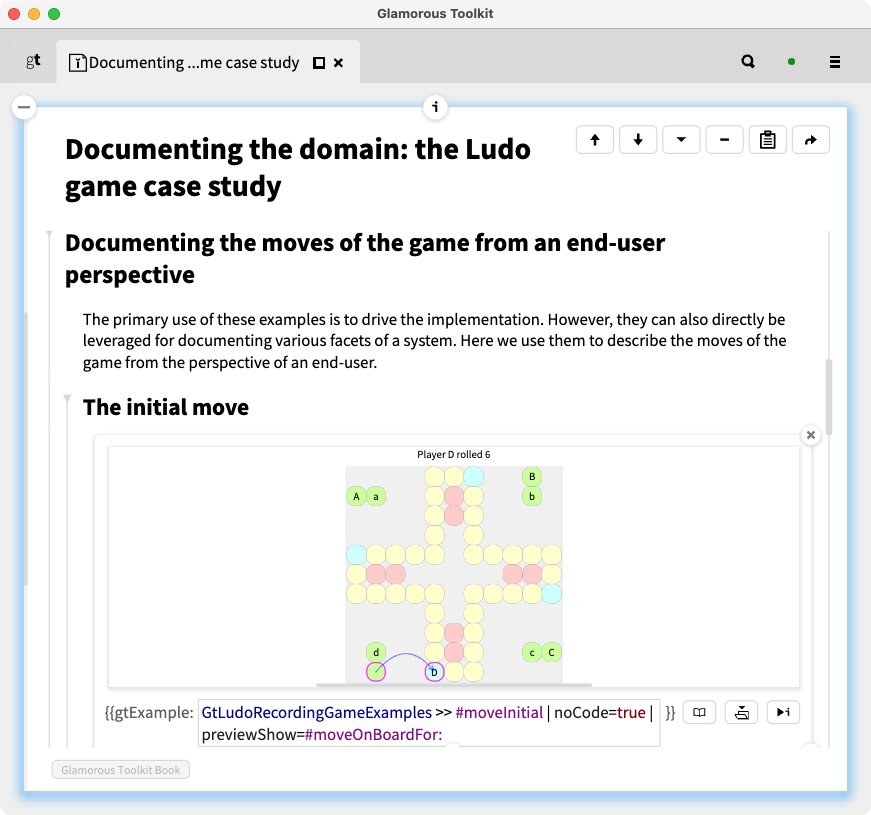}
  \caption{Documenting the Ludo game with live example objects.}
  \label{fig:documentingLudo}
\end{figure}

Examples such as these can be embedded into notebook pages and used as moldable objects for further development tasks, or as documentation, as seen in \autoref{fig:documentingLudo}.

\patsec{Consequences}
Examples can be run just like classical unit tests.\\
When an example fails, its dependent examples do not need to be run.\\
When an example succeeds, it can be inspected, used as a moldable object to start coding, or embedded as a live example snippet within a notebook page to illustrate some point.\\
When you are searching for usages of an API, not only do you find examples that illustrate the usage, but by running the example you obtain a live instance that you can explore.

\patsec{Known Uses}
Subtext~\cite{Edwa04a} supports \emph{example-centric programming} by placing live examples at the focus of the development process.
JExample~\cite{Kuhn08a} is a Java-based testing framework in which tests return examples.

\patsec{Related patterns}
An example can be used as a \patref{MoldableObject} to start a new exploration activity.
Examples can also be embeddded within a \patref{ProjectDiary} notebook page to illustrate a particular documentation point.

%: ----- Moldable Data Wrapper -------------------------
\pattern{Moldable Data Wrapper}{MoldableDataWrapper}

%\ws{JN: Moldable Data Wrapper and Moldable Collection Wrapper seem to overlap or be in conflict.}

\patsec{Context}
You are working in a domain with existing data that you want to turn into an explainable system.

\patsec{Problem}
\emph{How do you develop custom tools for existing data?}

\patsec{Forces}
\begin{itemize}[---]
\item When we explore data, we represent them using suitable low-level data structures.
\item Data structures (lists, dictionaries) reflect the representation of data, not their interpretation.
\item To analyze and explore data, we need higher-level views that reflect our understanding of the data.
\end{itemize}

\patsec{Solution}
\emph{Wrap each kind of data using a dedicated class reflecting the problem domain entity.}

\patsec{Steps}
As you explore the data, introduce custom tools (views \etc) to the domain class that reflect answers to questions you ask about the data.

\patsec{Examples}
First extract the data of interest.
This might be data sitting in your file system (for example, a CSV file), or data retrieved from a website.
%\cp{Web APIs and databases also come to mind -- this sentence addresses my previous comment on the physical location of the data.}
For example, here we retrieve a \st{Dictionary} representation of JSON data about the feenk GitHub organization:

\begin{code}
url := 'https://api.github.com/orgs/feenkcom'.
json := ZnClient new get: url.
dictionary := STON fromString: json.
\end{code}

%\eog{Would love to see the goal or motivation earlier for this kind of jazz.}

\begin{figure}[h]
  \includegraphics[width=\columnwidth]{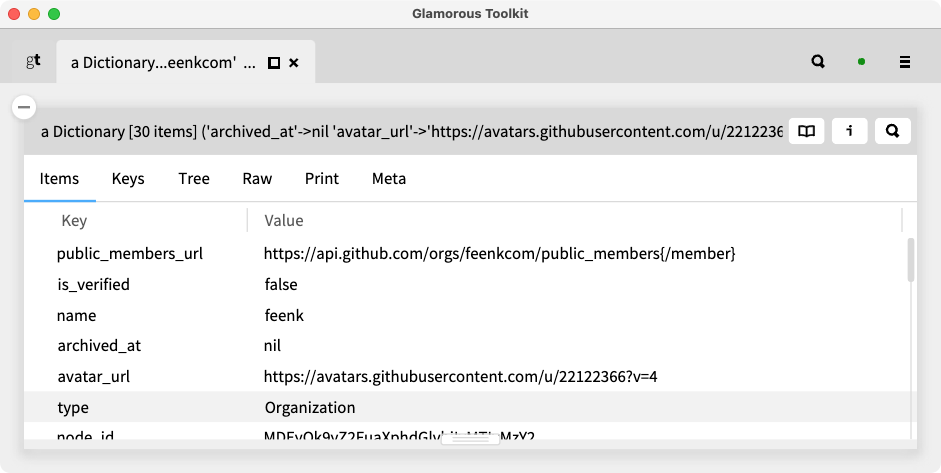}
  \caption{Raw (not moldable) GitHub data.}
  \label{fig:rawGitHubData}
\end{figure}

The dictionary representation, however, is not well-suited for exploring the GitHub organization domain.
If we explore the resulting object (see \autoref{fig:rawGitHubData}) we just see the keys and values of the raw downloaded data.
From this view, of course we can explore the data by navigating the Dictionary views, or by programatically exploring other paths, but we cannot add or tailor views to specifically support the GitHub Organization domain.

We address these problems by wrapping the raw data as a dedicated \st{GhOrganization} object, as follows:
\begin{code}
GhOrganization new rawData: dictionary.
\end{code}

\begin{figure}[h]
  \includegraphics[width=\columnwidth]{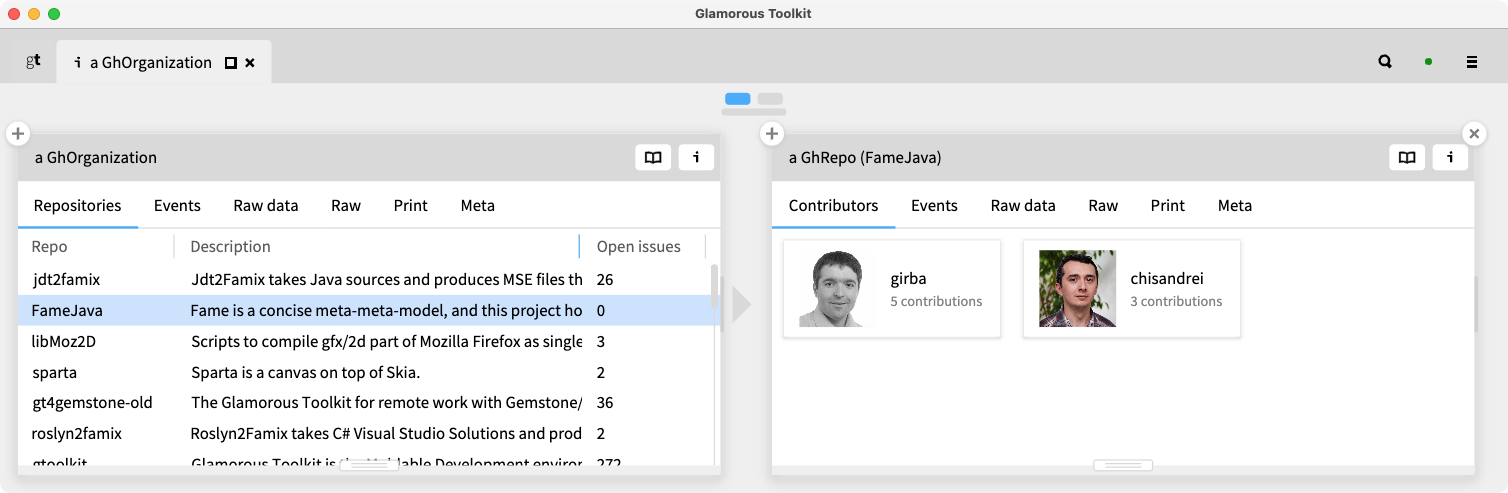}
  \caption{Wrapped (moldable) GitHub data.}
  \label{fig:wrappedGitHubData}
\end{figure}

Now we can add custom views specific to this domain, for example, listing the repositories of an organization, or the most recent GitHub events.
For each new domain concept, we introduce a dedicated wrapper object, so we can navigate the entire model via the domain concepts.
We see the result after some moldable development steps in \autoref{fig:wrappedGitHubData}.

\patsec{Consequences}
By wrapping the raw domain data, you obtain a moldable object that can be customized to form part of an explainable system.\\
Each time you navigate to other data representing further domain entities, wrap them as well, to build an explorable domain model.\\
In case custom tools of the underlying data objects are useful, you can always recycle them and make them available to the wrapped objects as well.

\patsec{Known Uses}
Lifeware\footnote{\url{https://www.lifeware.ch}} uses Moldable Data Wrappers extensively to wrap SQL data in the insurance domain to be able to create custom views for these data.
% 49 classes with gtViews and a rawData slot
Within GT (August, 2024) there are over 40 classes that wrap a \st{rawData} slot and provide custom views.
These classes wrap diverse data ranging from Jupyter notebooks to social media records.
EGAD~\cite{Vale23a} is a research prototype to explore GitHub actions by wrapping the raw data obtained from GitHub as moldable objects.

\patsec{Related patterns}
A \patref{CollectionWrapper} serves a similar purpose, but solves a different problem, which is to allow the results of queries to be moldable.

%: ----- Moldable Collection Wrapper -------------------------
\pattern{Moldable Collection Wrapper}{CollectionWrapper}

\patsec{Context}
Within your application domain, you not only have to deal with individual domain objects, but also composite entities (\eg a book of notebook pages, a website of web pages), and collections of entities (\eg the result of query).

\patsec{Problem}
\emph{How can you effectively provide custom tools for various kinds of collections of domain entities?}

\patsec{Forces}
\begin{itemize}[---]
\item Collections are generic data structures with generic views that are not necessarily informative for your domain.
\item Collections of domain entities may occur in various forms and contexts, such as the state of a composite object, or the result of a query.
\item Providing custom views for each of these can be tedious and lead to much duplicated code.
\end{itemize}

\patsec{Solution}
\emph{Wrap a collection of domain objects in a dedicated collection wrapper, and give it dedicated views, actions and searches.}

\patsec{Steps}
In case there are multiple composite entities or collections that should share the same custom tools, factor these out into a common abstract superclass or trait~\cite{Duca06b}.

\begin{figure}[h]
  \includegraphics[width=\columnwidth]{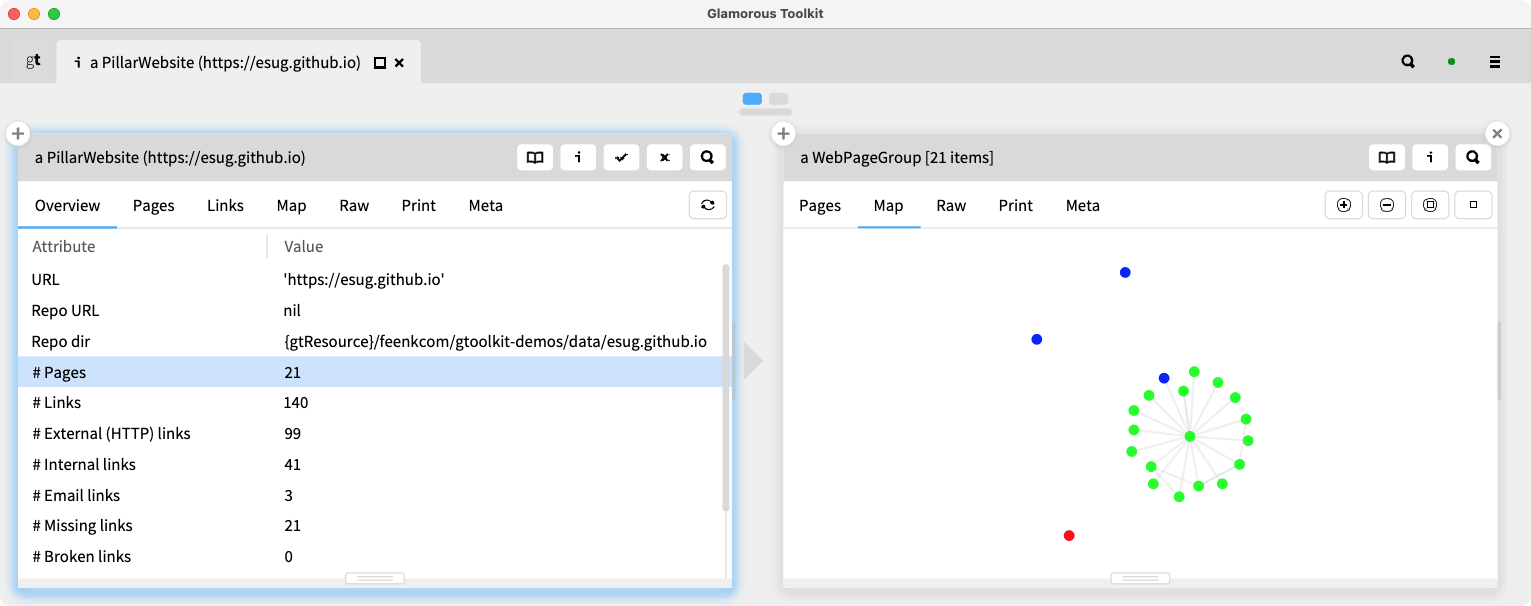}
  \caption{A moldable collection wrapper for web pages.}
  \label{fig:moldableCollectionWrapper}
\end{figure}

\patsec{Examples}
In \autoref{fig:moldableCollectionWrapper} we see that when we navigate to the \st{Pages} of a website, we obtain not a raw collection, but a \lst{Web\-Page\-Group} wrapping the collection, and providing custom tools, such as a map of reachable and unreachable pages.
Similarly, a query for pages matching a particular string will also return a wrapped group with dedicated tools.

Furthermore, if the wrapped collection or composite object should also behave like a collection, use a trait to forward collection API requests to the underlying collection.
In this example, \st{WebPageGroup} uses the trait \st{TGtGroupWithItems} that forwards all collection requests to the \st{items} slot of the wrapper that holds the raw collection.

\patsec{Consequences}
Composite domain objects and collections of domain objects can be molded to support custom tools.
Query results can be similarly molded.\\
Wrapped collections can be made plug-compatible with raw collections.

\patsec{Known Uses}
Lifeware\footnote{\url{https://www.lifeware.ch}} uses Moldable Collection Wrappers to wrap SQL query results so the query results can be molded.
There are about two dozen usages with GT, wrapping collections of notebook pages, web pages, logging events, and various other kinds of domain entities.

\patsec{Related patterns}
Whenever you develop a \patref{CustomSearch} that returns a collection, consider wrapping it as a Moldable Collection Wrapper.

% ===== Process =========================
\section{Process Patterns}\label{sec:process}

We conclude with the process patterns that provide guidance in organizing and steering moldable development tasks.

%: ----- Project Diary -------------------------
\pattern{Project Diary}{ProjectDiary}

\patsec{Context}
You are working on a software project and need to track your progress.

\patsec{Problem}
\emph{How can you keep track of decisions, experiments and progress in a moldable development project?}
%\cp{consider mentioning this property in the context already (unless the pattern is applicable to any software project)}

\patsec{Forces}
\begin{itemize}[---]
\item It's boring to write documentation after the fact.
\item Documentation is not part of the running system, so it distracts you from coding.
\item Tools for tracking your progress are separate from the code base, so they are commonly out of sync with reality.
\end{itemize}

\patsec{Solution}
\emph{Use a live notebook 
%\cp{why "live"? and isn't this notebook a place to write some kind of documentation?}
as the starting point for all tasks.}

\patsec{Steps}
Create a dedicated notebook page for each project, or even each project task, to summarize the goals, and provide pointers to related material.
Use the notebook to keep a diary of your progress.
As the project matures, use the notebook as a draft 
%\cp{will the "draft" be eventually discarded or become permanent documentation at some point? and how to avoid it looses "sync with reality"?}
for the documentation.

\begin{figure}[h]
  \includegraphics[width=\columnwidth]{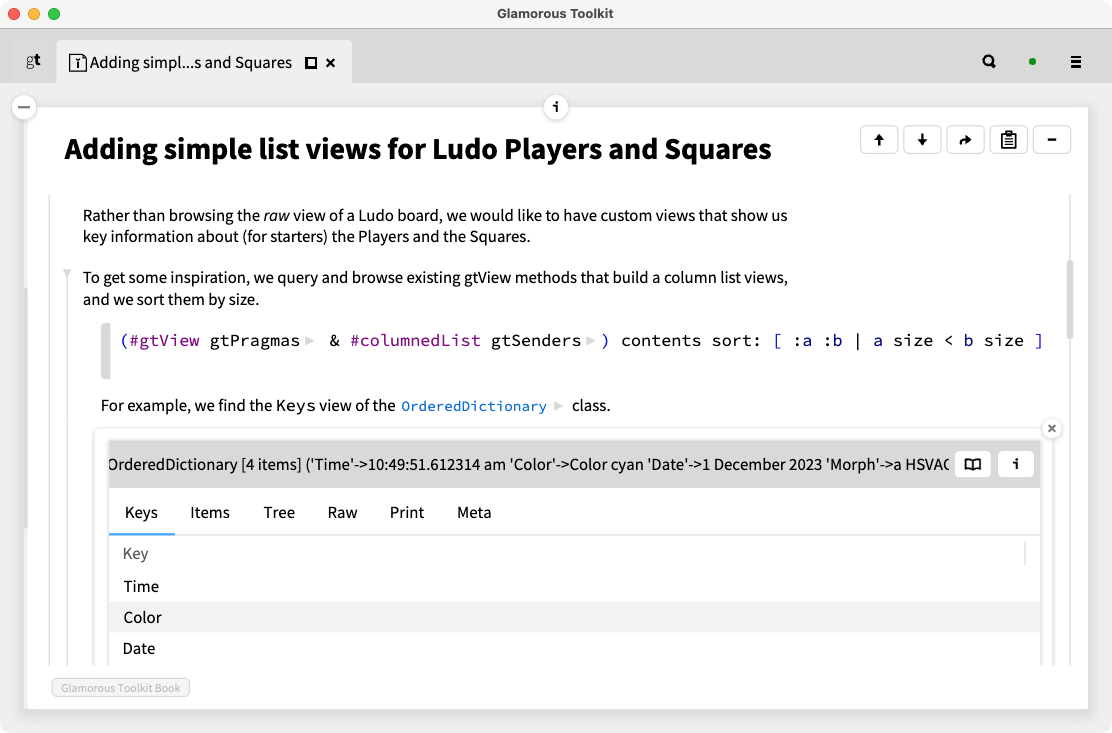}
  \caption{Documenting your progress in a live notebook.}
  \label{fig:ProjectDiary}
\end{figure}

\patsec{Examples}
In \autoref{fig:ProjectDiary} we see a live Lepiter page in the \GT Book
%\footnote{An archived static snapshot of this page is also available: \url{https://web.archive.org/web/20231213144857/https://book.gtoolkit.com/adding-simple-list-views-for-ludo-players--3g9kpmxty3fj9eev3ucmqkztc}.} 
documenting the task of adding some list views to the Ludo game.
This page started as a Project Diary page, and was later rewritten as documentation.

%\eog{``Using a live notebook.'' I know folks maybe in like Jupiter like Python stuff but specific kind of like scientific Python are used to using this word or this phrase, but I would explain it or define it if you can, if you if you have the space.}

If needed, add code snippets for any setup tasks (\eg cloning repositories or loading databases).
Also add code snippets to the notebook pages to serve as moldable objects to start coding from.
Extract interesting code snippets as example objects to document interesting use cases, or to serve as tests.

As the project grows, organize the notebook into a main page with an overview, and separate notebook pages for different tasks or groups of related tasks.

Consider using notebook tags 
%\cp{related work?: \url{https://hillside.net/plop/2013/papers/proceedings/papers/cunningham.pdf}}
to organize your pages implicitly. For example, use a dedicated project tag for all the project pages, and additional tags to indicate their status (``todo'', ``completed'', ``urgent'' \etc).
For example, the ``DRAFT'' tag is used to track pages of the \GT Book that require further editing.

At the end of a project, consider recycling and rewriting the project pages to create documentation. In this way the diary can serve as a rough draft.

% \kh{The "Project Diary" pattern is a very useful one, and also heavily used in other contexts (in science we call it a "lab notebook"). But there is one aspect of it to which is more an ideal than reality: "At the end of a project, consider recycling and rewriting the project pages to create documentation. In this way the diary can serve as a rough draft." In my experience, this rarely happens, and Lepiter supports it even less than other tools. Typically you want to keep the diary for yourself as a historical record, so morphing it into documentation for others by successive edits is not a good approach. Ideally, I'd want to be able to mark parts of it for inclusion in a new document that I'd then edit, but I haven't yet seen any support for this. In practice, documentation gets written from scratch, with the diary serving only as a source of detailed information to be copy-pasted. Maybe one day we will see refactoring tools for prose.} 

\patsec{Consequences}
A live notebook is an integral part of the project, and evolves together with other software artifacts.\\
Notebook pages can express tasks in various stages of completion, so can be used as starting points for further development tasks.\\
Notebooks can serve not only to track progress, but also as documentation for both technical and non-technical stakeholders.

\patsec{Known Uses}
A Zettelkasten\footnote{\href{https://web.archive.org/web/20240917092258/https://en.wikipedia.org/wiki/Zettelkasten}{https://en.wikipedia.org/wiki/Zettelkasten}}~\cite{Luhm81a} is a traditional form of note-taking, where notes are cross-indexed in card files using various metadata, to be used as a knowledge base for research and writing.
Jupyter\footnote{\href{https://web.archive.org/web/20240820082136/https://docs.jupyter.org/en/latest/}{https://docs.jupyter.org/}} notebooks are commonly used as both activity logs and as project documentation.
Personal knowledge management systems like RoamResearch\footnote{\href{https://web.archive.org/web/20240820082136/https://docs.jupyter.org/en/latest/}{https://roamresearch.com}} are regularly used for diaries for software projects.

\patsec{Related patterns}
A Project Diary page can embed numerous live \patref{ExampleObject} instances to document specific points.

%: ----- Tooling Buildup -------------------------
\pattern{Tooling Buildup}{ToolingBuildup}

%\kh{A close second (of a non-pattern) is "Tooling Buildup". Has anyone outside feenk ever done this, beyond tweaking some parser? These two (non-) patterns are probably the biggest obstacle to getting started with Moldable Development. If your domain or your tech stack is not supported by GT out of the box, testing the idea is a very costly endeavor.} 

\patsec{Context}
You want to start moldable development but lack the basic infrastructure 
%\cp{is the IDE part of this basic infrastructure? what are the assumptions about the IDE itself?}
to start exploring the domain.
You may be missing a parser for a data format or a DSL, a language bridge to an execution platform, or implementations of key analysis algorithms. 

\patsec{Problem}
\emph{How do you start moldable development if you are missing core infrastructure?}

\patsec{Forces}
\begin{itemize}[---]
\item When you are building on top of an existing project, you may be missing key tools to allow you to start exploring the domain.
%\dd{This feels like more of the problem.}
\item You might not know where to find existing implementations to directly use, or to port or adapt to your development platform.
\item You might lack the expertise to build the tools yourself.
\end{itemize}

\patsec{Solution}
\emph{To start a moldable development activity, first obtain the basic tools you need to start exploring.}
%If these are missing, you need to first work on a \emph{tooling buildup} phase.

\patsec{Examples}
In this phase you \emph{do not focus} on specific domain questions you want to explore, but rather on acquiring the tools you need to start exploring:
\begin{inparaenum}[(i)]
\item building a parser for YAML configuration files so that you can extract useful data from them and apply the \patref{MoldableDataWrapper} pattern,
\item building an island parser~\cite{Kurs14b} to extract interesting bits of information from source files 
%\cp{sounds like a form of static analysis?}
in software languages for which a full parser may not be readily available,
\item building a bridge to a foreign execution platform, such as for Python, AWS, or a database system, so that you can observe and interact with run-time entities in the target platform, or
\item implementing graph-traversal algorithms to detect dependencies, deadlocks \etc
\end{inparaenum}

\patsec{Consequences}
Tooling Buildup will slow you down.\\
%\cp{especially if the expertise is lacking. Most of the previous solution examples imply something needs to be built. However there are YAML parser libraries, SQL database clients, etc. Is reusing them vs. rewriting them from scratch not also part of "tooling buildup"?}
Once you have the right tools in place, you can move fast.
%\dd{Good use of opposing consequences!}
%\ws{DP: In Tooling Buildup has a negative consequence. But they are predominantly positive. Need to see more potential liabilities.}

\patsec{Known Uses}
Tooling Buildup is arguably a common pattern within just about any kind of project.
We mention it explicitly here to emphasize the point that moldable development does not necessarily work out-of-the-box, but may require some up-front investment.

\patsec{Related patterns}
A \patref{MoldableTool} is part of the moldable development environment, whereas Tooling Buildup is about the additional, specific tools you need before you can start moldable development.

%: ----- Blind Spot -------------------------
\pattern{Blind Spot}{BlindSpot}
\patsec{Context}
You are starting to work with an existing team and code base, and you need to engage the key stakeholders.

\patsec{Problem}
\emph{How do you pick the first moldable development task to focus on?}

\patsec{Forces}
\begin{itemize}[---]
\item The stakeholders are already heavily committed to their current development tasks, and have little time to spare for you.
\item They are almost certainly skeptical that moldable development will make them more productive.\\
You need to demonstrate early success to get the customer to commit long-term to further collaboration.
\item It can be difficult to pick a task that is both feasible in a short amount of time, and also brings value to the stakeholders.
\end{itemize}

\patsec{Solution}
\emph{Find the parts of the system that are creating problems for the stakeholders and make them explainable.}

\patsec{Steps}
The ``blind spots'' are the aspects that are difficult to understand, to monitor, or to debug.
Pick one of these blind spots as a target.
The task should be feasible in reasonable time
(\ie hours or days, but not weeks or months)
%\cp{such as minutes, hours, days, weeks?}
but non-trivial.
There should be a clear value for the stakeholder, that is, do something that the stakeholder has difficulty with.

%\eog{I really like this blind spot pattern. pointing out things that are difficult to understand modern debug. I almost want to like, say this is another like first place that we should talk about. You do that. I just wonder like should we like lift this further up in the paper or talk about it first?}

\patsec{Examples}
Some typical examples are 
\begin{inparaenum}[(i)]
\item exposing hidden dependencies between features, 
\item visualizing performance costs of test runs on a cluster, or
\item providing views to gain insights into scheduling deadlocks.
\end{inparaenum}

\patsec{Consequences}
Since 
%\cp{Sounds already like a consequence (?)}
the stakeholder has little time to spare to guide you, you may have to work in the dark at times.
Be sure to always have something to show when you come back with questions.\\
By quickly developing a solution to an ongoing problem, you will engage the stakeholder.\\
It does not matter if the custom tools you develop do not perfectly match the stakeholder's needs, as the only goal is to convince the stakeholders to commit to closer and longer-term collaboration.

\patsec{Known Uses}
%\tg{This is an example from a 2-day hackathon: \url{https://lepiter.io/feenk/steering-agile-architecture-by-example--th-e2p6aps2brbby94deek31xqxh/}. However, while it worked nicely, it did not convince anyone at that time.}
%\todo{Need some GT anecdotes. Can we expand on one of the examples listed above?}
Finding and understanding legacy feature toggles in the Open edX Proposals project is a public example of a Blind Spot tackled with GT during a hackathon~\cite{Girb21b}.

\patsec{Related patterns}
A Blind Spot will manifest itself as one or more instances of a \patref{MoldableObject}.

%: ----- Simple View -------------------------
\pattern{Simple View}{SimpleView}

%\ws{JN: Instead of Simple View, why not just take a generic view? What's the difference between a simple view and a custom view?}
%\todo{Explain the difference between a Simple View and a raw view.}

\patsec{Context}
You have found some useful domain information to expose and have decided to implement a \patref{CustomView}.

\patsec{Problem}
\emph{How should you design a custom view to effectively communicate the information of interest?}

\patsec{Forces}
\begin{itemize}[---]
\item There are many different ways to design a view.
\item It is hard to anticipate what information another Stakeholder would be interested in seeing.
\item You can spend an arbitrary amount of time and effort implementing a rich, interactive visualization as a view.
\end{itemize}

\patsec{Solution}
\emph{Always start with the simplest kind of view that you can implement quickly to convey just the information you are interested in for your current task.}

\patsec{Steps}
Enhance the simple view, or replace it by a richer view only when there is a clear need to do so.

This pattern follows exactly the same reasoning as the \emph{Goal/Question/Metric} (GQM) approach~\cite{Basi94a} for designing software metrics.
A view, like a metric, only exists because it helps you to answer a question that supports you in achieving some goal.
Views or metrics that do not support a specific goal or answer a concrete question have no reason to exist.
It follows, then, that a view should be designed to answer a specific question you have in a development task.
The simplest design that achieves this is the one you want.
%\cp{achieving the peak of simplicity may also require to spend an arbitrary amount of time and effort :-)}
If new goals and questions arise, you will then either design new views, or extend or replace the existing views by richer ones.

\begin{figure}[h]
  \includegraphics[width=\columnwidth]{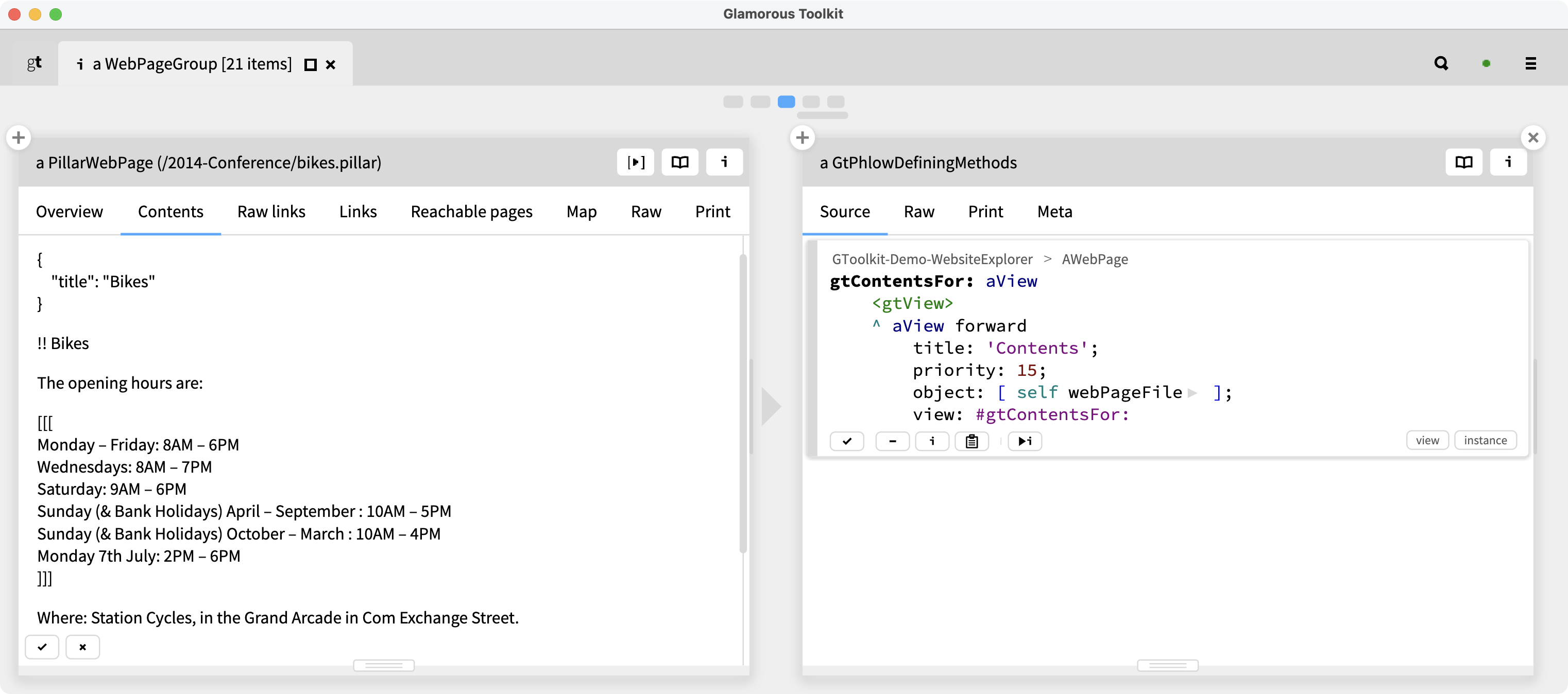}
  \caption{A webpage contents view forwarding to its file's contents view.}
  \label{fig:forwardView}
\end{figure}

\patsec{Examples}
\GT supports a wide range of different kinds of views, ranging from simple textual lists and trees, to arbitrary visualizations.
However, the most widely used view in the default environment is the \st{forward} view, which allows an object to reuse a view of a component or a collaborator.
We can see this in \autoref{fig:forwardView}, where a webpage object leverages the existing \st{gtContentsFor:} view of the underlying file to display the file's contents.
This is also the simplest view to implement, since it only requires, at a minimum, the specification of the target object (\st{self webPageFile}) and the view method to reuse (\st{#gtContentsFor:}).

The next most popular view is a \st{columnedList}, which presents a collection of data in a format resembling a spreadsheet with column headers.
The \st{Moves} view in the middle pane of \autoref{fig:rawViewVsCustomView} is an example.
A common flow is to start with a \st{forward} view to an existing \st{columnedList} view of another object, then replace it by a custom \st{columnedList} when a need is identified to rename, replace or add columns.

%\eog{The forward view and the column list.
%I think talking about the forward view with just this is hard to digest.
%Maybe a picture here would help with other kinds of simple views.}

NB: A \emph{raw view} (leftmost pane of \autoref{fig:rawViewVsCustomView}) is \emph{not} a simple view.
Although it is ``simple'' in the sense that it requires no effort to implement, it does not succeed in focusing your attention to the information of interest.
With a raw view, it is almost always necessary to navigate through the view to find what you want.
Although the Raw view and the Moves view are showing us the same information in \autoref{fig:rawViewVsCustomView}, only the the (simple) Moves view highlights this information in a clear and accessible way.

Similarly, the \st{Board} view of the Ludo game (leftmost pane of \autoref{fig:ludoViews}) is \emph{not} a simple view, as it requires some effort to build up the interactive GUI view of the board.
%\cp{this sounds more like a domain-specific view. The views mentioned before are rather generic and reusable (but not necessarily simple)}
On the other hand, once we have such a view, it can be reused and repurposed, as we can see in the \st{Move} view of the Ludo Move object in the same figure.

Other kinds of simple views exist to display plain text, simple lists without columns, and trees of hierarchical data.

\patsec{Consequences}
A simple view requires little effort to implement and can be quickly deployed and tested.\\
Simple views are easy to understand and extend or replace.

\patsec{Known Uses}
Notebooks like Jupyter make it easy to generate simple views as interactive visualizations.\footnote{\href{https://web.archive.org/web/20240430071604/https://jupyterbook.org/en/stable/interactive/interactive.html}{https://jupyterbook.org/en/stable/interactive/interactive.html}}
Within GT there exist nearly 1000 \st{forward} views that simply reuse an existing view of another object.
Over 800 views are simple \st{columnedList} views, in contrast to just under 500 \st{explicit} views that explicitly construct more complex graphical views.

\patsec{Related patterns}
When you introduce a \patref{CustomView}, you want to start with a Simple View.

%: ----- Throwaway Analysis Tool -------------------------
\pattern{Throwaway Analysis Tool}{ThrowawayAnalysisTool}

%\ws{JN: Throwaway Tools aren't.}

\patsec{Context}
You are assessing a problem such as a bug, a performance problem, or a security issue, and you are facing a problem that is very difficult to understand.

\patsec{Problem}
\emph{How can you advance the assessment effectively?}

\patsec{Forces}
\begin{itemize}[---]
\item The goal of an assessment is to find answers as quickly as possible.
\item Custom tools help make analyses faster, while generic, ``reusable'' tools can make it more difficult to find the answers you need.
%\cp{building generic tools is definitely more expensive than building ad-hoc tools. But the "reusable" tools by definition are already built and just need to be selected or rejected as not suitable for the task, a decision that shouldn't take too much time. Clearly if the generic tool has been selected for performing a task for which it is not appropriate...}\\
\item Throwaway tools aren't.
\end{itemize}

\patsec{Solution}
\emph{Build a dedicated, throwaway tool just for your problem.}

\patsec{Steps}
Use a \patref{ContextualPlayground} to prototype a visualization or to explore the result of a quick query.

The focus should be on finding an answer.
Reusability should only later become a concern.
%\cp{this discussion on reusability/generality vs. ad-hoc/specificity belongs more in the consequence part vs giving details on the solution on how to rapidly build a specific tool dedicated to the problem}
Some tools can indeed become reusable as a \patref{CustomView}, \patref{CustomSearch}, or \patref{CustomAction}, but that should not be the driving force.
When you start with reusability in mind, you tend to favor generic tools. 
%\cp{is this about bulding a tool or selecting a tool?}
However, assessment benefits from \emph{contextual tools}.
Focusing on the context first speeds up the overall effort. 
To make this economically feasible, the cost of the tool should be amortized on the first use.
%\cp{which implies that such cost should be very low -- making the "building" vs "using" (not "reusing") a tool concept intriguing}
You can make this possible through a \patref{MoldableTool}.

\begin{figure}[h]
  \includegraphics[width=\columnwidth]{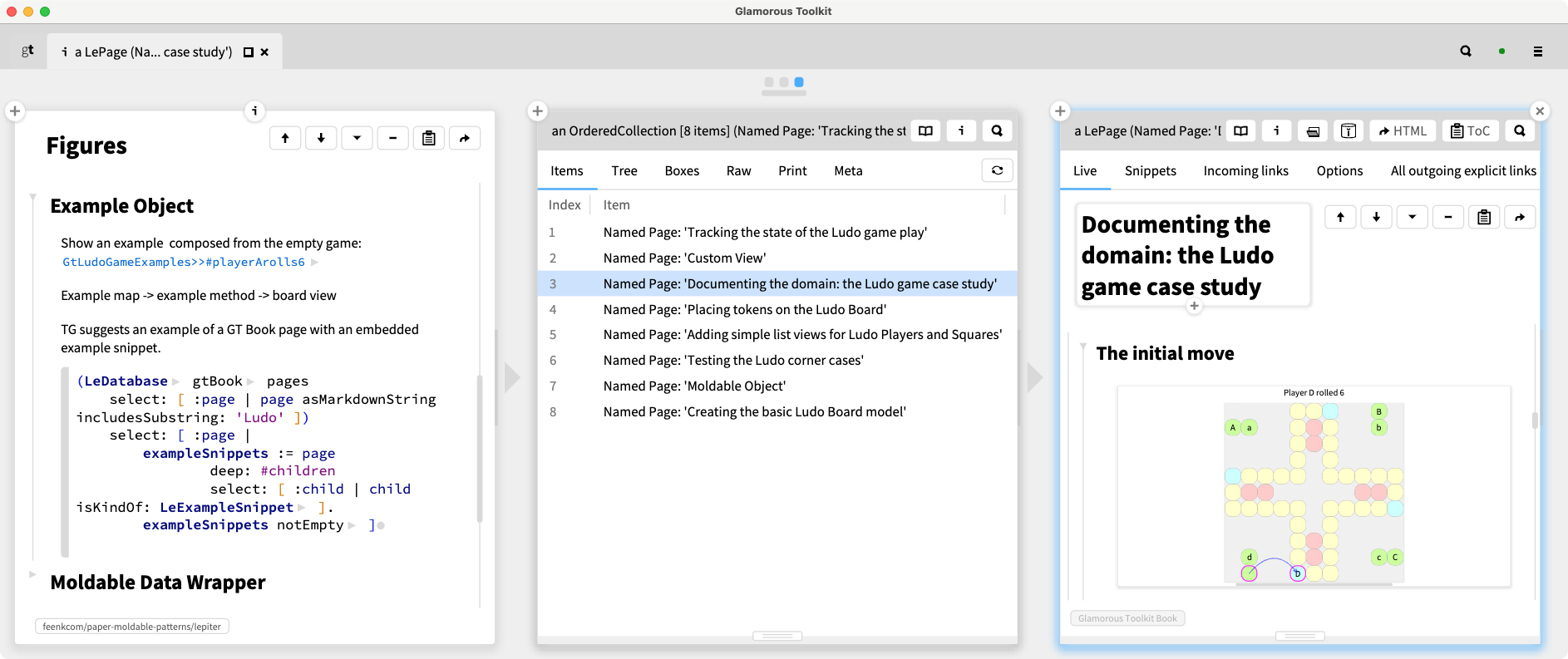}
  \caption{A throwaway analysis to find candidate examples.}
  \label{fig:throwawayQueryTool}
\end{figure}

%\eog{Project Diary, I think this is really cool.
%It's just not legible in this size.
%I wonder if splitting it over to figures to enhance its legibility.}

\patsec{Examples}
A throwaway analysis tool should be cheap to implement.
%\cp{consider moving closer to the "economically feasible" sentence above}
In \autoref{fig:throwawayQueryTool} we see an \emph{ad hoc} query to find Lepiter \patref{ProjectDiary} pages that match the string ``\st{Ludo}'' and also contain \patref{ExampleObject} snippets, to illustrate the use of example objects in documentation.

Throwaway analysis tools may appear to be wasteful: they cost development effort that you do not get to reuse.
However, the goal is to optimize the overall development speed.
The alternative to custom tools is manual exploration.
Building a tool, even for just one usage, can outcompete manual exploration by a large margin.
That budget difference can more than make up for the cost of the tool development.
%\cp{obligatory https://xkcd.com/1319/}

\patsec{Consequences}
By making it cheap to build analysis tools, 
%\cp{the solution above should highlight in more detail how to actually lower the cost of such tool building}
tool-building becomes an essential part of the development process, rather than a side activity, just as writing tests has become an integral development activity rather than an add-on.

\begin{figure}[ht]
  \includegraphics[width=\columnwidth]{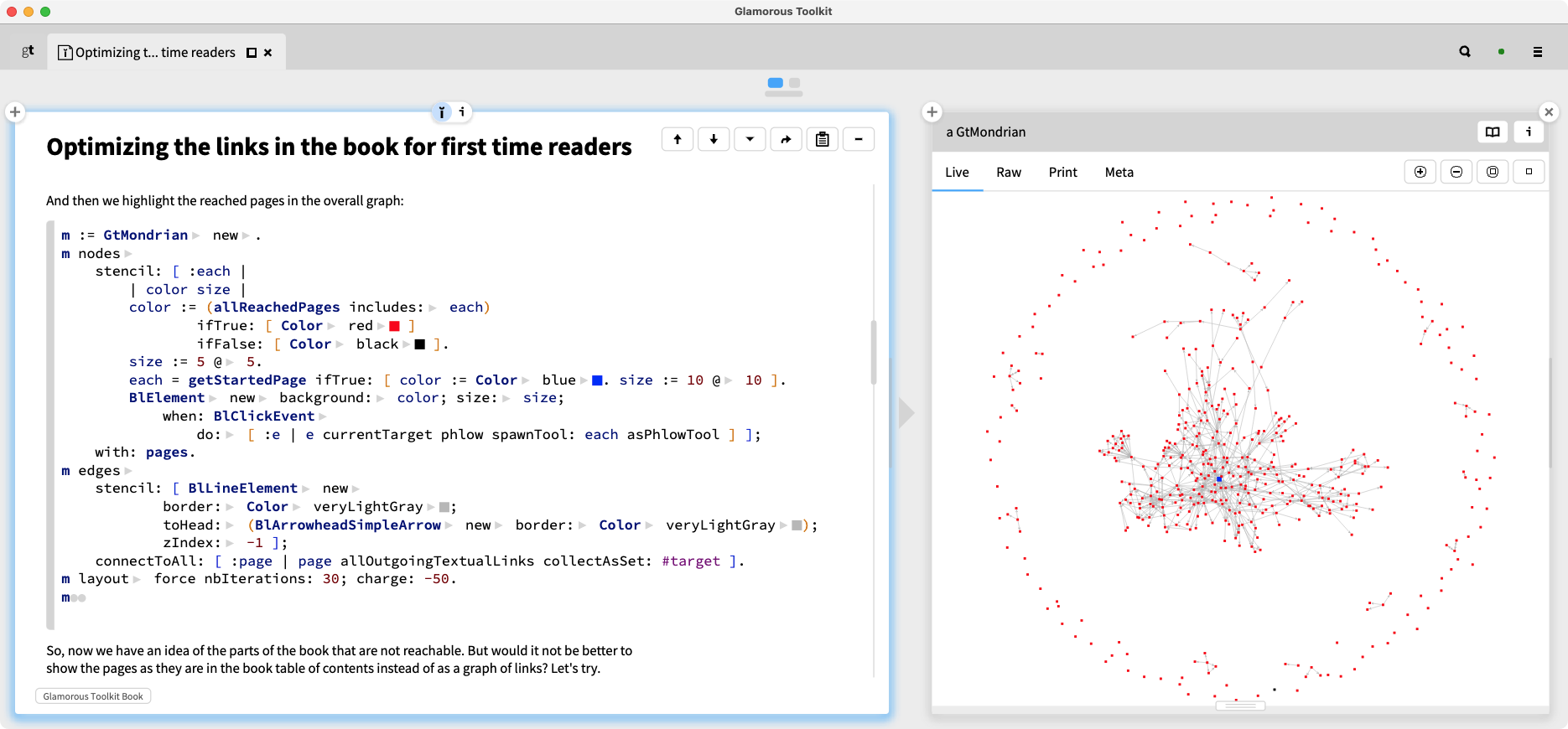}
  \caption{A throwaway tool to visualize reachable and unreachable knowledge base pages.}
  \label{fig:optimizingLinks}
\end{figure}

\patsec{Known Uses}
There are numerous examples of \emph{ad hoc} analyses implemented as Throwaway Tools in the GT book, for example the page on \emph{Optimizing the links in the book for first time readers} shows a script (\autoref{fig:optimizingLinks}) to generate a visualization of the pages in the book itself that are reachable by following links contrasted with standalone pages.

\patsec{Related patterns}
Use a \patref{ContextualPlayground} to prototype the tools.

% ===== Conclusion =========================
\section{Conclusion}

%\ws{JN: The voice of the author is disembodied, but only comes out in the conclusion that there is a lot of experience here. It should be made clearer earlier.}

The pattern language described here has been mined from several years of experience in applying moldable development to both open- and closed-source projects, using the \GT moldable development platform.
Moldable development makes software systems explainable by making it cheap to add custom tools that support decision making.
In practice, however, this requires a paradigm shift since it leverages live programming to explore domain objects and incrementally mold them.

Although the examples given here have all been drawn from \GT, we believe that they can be applied in any programming environment that offers a minimal level of support for live programming.
The key to supporting moldable development is the pattern \patref{MoldableTool}.
The tools of the environment need to be open to small, custom changes that support decision-making by answering questions about the underlying system under development.
An example of another system that supports such changes is Emacs~\cite{Stal81a}, though it was not designed to support moldable development as we describe it.
Once the development environment has been opened up to make its core tools moldable, then moldable development can truly start.

% ===== Acknowledgements =========================
\subsection*{Acknowledgements}
We would like to thank Ralf Barkow, Ward Cunningham, Konrad Hinsen, Timo Kehrer and Edward Ocampo-Gooding for their detailed reviews of drafts of this paper.
% Alberto Torres, -- didn't reply
Thanks are due as well to Petr Picha, Cesare Pautasso, and Daniel Pinho for their written comments, and the EuroPLoP workshop participants for their spoken feedback.

% ===== References =========================
\ifthenelse{\boolean{preprint}}{% FOR PREPRINT
\renewcommand*{\bibfont}{\small}
}{}
\bibliographystyle{ACM-Reference-Format}
\bibliography{moldablePatterns}

\end{document}